\documentclass[prb,showpacs,twocolumn,superscriptaddress]{revtex4}
\usepackage{hyperref,graphicx,dcolumn}
\usepackage{amsfonts,amsmath,amssymb,bm}
\usepackage{color}
\usepackage{pdfsync}

\newcommand{\epsdir}{./graphs}

\newcommand{\myFig}[6]{ %
\begin{figure}[htb] 
\begin{center} 
\includegraphics[width=#1\columnwidth,height=#2\columnwidth,clip=true,keepaspectratio=#3]{\epsdir/#4}
\caption{#5} \vspace{-0.5cm} \label{#6} 
\end{center} \end{figure}} 

\newcommand{\vcr}[1]{\boldsymbol{\mathrm{#1}}}

\newcommand{\braket}[1]{\left\langle #1 \right\rangle}

\newcommand{\Abs}[1]{\left\vert #1 \right\vert}

\newcommand{\SqB}[1]{ \left[ #1 \right]}
\newcommand{\RnB}[1]{ \left( #1 \right)}
\newcommand{\CrB}[1]{ \left\{ #1 \right\}}
\newcommand{\cool}[1]{\mathcal{#1}}
\newcommand{\upup}{\uparrow\uparrow}
\newcommand{\updo}{\uparrow\downarrow}

\begin{document}
\title{Dynamical exchange interaction from time-dependent spin density functional theory}
\author{Maria Stamenova}\email[Contact email address: ]{stamenom@tcd.ie}
\affiliation{School of Physics, Trinity College, Dublin 2, Ireland} 
\author{Stefano Sanvito} 
\affiliation{School of Physics, Trinity College, Dublin 2, Ireland} 

\begin{abstract}
We report on {\it ab initio} time-dependent spin dynamics simulations for a two-center magnetic molecular complex based on time-dependent non-collinear spin density functional theory. In particular, we discuss how the dynamical behavior of the {\it ab initio} spin-density in the time-domain can be mapped onto a model Hamiltonian based on the classical Heisenberg spin-spin interaction $J\vcr{S}_1\cdot \vcr{S}_2$. By analyzing individual localized-spin trajectories, extracted from the spin-density evolution, we demonstrate a novel method for evaluating the effective Heisenberg exchange coupling constant, $J$, from first principles simulations. We find that $J$, extracted in such a new dynamical way, agrees quantitatively to that calculated by the standard density functional theory broken-symmetry scheme. 
\end{abstract}


\maketitle
\section {Introduction}

In magnetic recording the typical time-scale for magnetization reversal is in the nanosecond range, and it is now believed that the ultimate limit for magnetization switching by magnetic field pulses may approach the picosecond mark~\cite{Tudosa2004}. Down to the picosecond scale, the exchange interaction is constant in time and so is the magnetic anisotropy. This allows the dynamics of magnetization to be modeled in terms of the continuous Landau-Lifshitz-Gilbert equation~\cite{LLG}, usually solved with micromagnetic techniques \cite{Miltat}. The spatial resolution of such techniques is chosen in view of the problem at hand and numerical considerations, but the equation of motion is always the same. At the most fine-grained and microscopic end of the modeling spectrum there are atomistic spin models, which have been proved to be a powerful tool for approaching the extreme phenomenology of the {\it ultrafast} magnetization dynamics \cite{Atxitia2009}. In these one associates classical spin-vectors to magnetic atoms, $\vcr{S}_i$, which are then coupled through a time-independent Heisenberg Hamiltonian 
\begin{equation} \label{Jij}
H = - \sum_{i>j} J_{ij} {\vcr{S}}_i \cdot {\vcr{S}}_j \,,
\end{equation}
where $J_{ij}$ are the pairwise Heisenberg exchange parameters. The state of the art for the theory is then represented by performing atomistic dynamical simulations, in which the Heisenberg Hamiltonian is completed by various spin-orbit terms, by the interaction with an external magnetic field and eventually by stochastic fields~\cite{Atxitia2009,Skubic2008}. 

The parameters of the theory, the exchange integrals and the anisotropy, are usually fitted to experiments or calculated from static density functional theory (DFT)~\cite{Skubic2008}. In this second case usually the exchange is obtained with the, so called, {\it broken symmetry} approach, proposed first by Noodleman\cite{Noodleman}. In its DFT variant broken symmetry refers to an unrestricted-spin calculation for open-shell complexes, where opposite spin densities are allowed to localize at different atomic sites. This broken symmetry or low spin (LS) state, unlike the state with the highest spin (HS), is not an eigenstate of the full spin operator (hence the name). The exchange parameters are then determined as differences between the total energy, $E_\alpha$ ($\alpha=$~LS, HS), of the different spin state
\begin{equation} \label{JBS}
J_{ij} = f\RnB{S_i,S_j} (E_\mathrm{LS} - E_\mathrm{HS})\:,
\end{equation}
where different formulations of the spin-dependent function $f$ are possible, depending on the choice of basis and level of localization, and where $S_{i}$ is the expectation values of the local spin at atom $i$ (see for instance references [\onlinecite{Ruiz,Peralta}]). 

The first demonstration of laser-induced ultra-fast demagnetization~\cite{Bea1996} in transition metals, however, opened a new frontier, namely, the possibility of manipulating and controlling the magnetization with ultrashort intense laser pulses~\cite{Kiril2010}. Here one reaches the femtosecond time resolution, where both the exchange interaction and the anisotropy may become time-dependent. Most importantly, in this limit the approximation of associating a classical spin of constant magnitude to an atom may breakdown. It makes sense that at a time-scale where the electronic degrees of freedom evolve in time in a non-adiabatic way (the local magnetic moment changes in time), spin-dynamics needs to be addressed at the electronic level. Yet, in order to interpret the results in a simple and transparent way, it is desirable to be able to map the electronic time-dependent simulations onto classical atomic models based on the Heisenberg Hamiltonian. How to perform such mapping, and whether this is at all possible, is the subject of the present paper. In particular, we will discuss how the evolution in time of the spin-density in time-dependent DFT~\cite{TDDFT} (TDDFT, or, to be more specific, its extension to non-collinear spin~\cite{Sandratskii}, the TDSDFT~\cite{TDSDFT}) simulations can be used to extract an effective spin-dynamics, which in turns can be mapped on an Heisenberg Hamiltonian. As a byproduct of such analysis we will be able to extract exchange parameters, whose values are quantitatively rather close to those calculated with the broken symmetry approach.

The paper is organized as follows. In the next section we will discuss the most technical aspects of our work. In particular, we will present the classical solution of the time-dependent Heisenberg model for a diatomic molecule. This will be useful to interpret our TDDFT results. In the same section we will describe the general aspect of the TDDFT simulations and explain how to integrate the charge density in order to map the TDDFT results onto the Heisenberg model. Then, in the following two sections, we will present results for both a stretched H$_2$ dimer and a hypothetical H-He-H trimer. These are qualitatively different systems with respect to the spin-spin interaction. In H$_2$ the spins of the two H atoms are coupled via direct exchange, while in H-He-H the exchange is indirect, {\it superexchange}~\cite{Yosida}, across the closed shell He atom. Finally we will conclude.

\section{Magnetic dimer: Theoretical aspects}

\subsection{Implementation of the TDDFT method}

{\it Ab initio} spin dynamics is simulated in the time-domain with the state of the art TDDFT code {\sc Octopus}~\cite{oct1}. This is a open-source (GPL) package capable of simulating excitations of molecules or clusters to custom-designed electromagnetic fields beyond the linear response regime, i.e. by the explicit time-propagation of the TD Kohn-Sham equations in a basis-free real-space representation. {\sc Octopus} provides an ideal environment for examining fundamental processes in the time-domain. Our starting point to understand {\it ab initio} spin dynamics in the time-domain has to be through the simplest complexes of non-spin-singlet atoms. In fact, the simplest possible real system for which the Heisenberg spin Hamiltonian of Eq.~(\ref{Jij}) was originally conceived, is the hydrogen molecule. In its ground state H$_2$ is closed-shell (diamagnetic), but in the stretched dissociating state local spins can be defined for each of the hydrogen atoms (e.g. in the broken symmetry LS state, the electrons localized on the opposite protons have particular and opposite spin expectation value, $s^z_{1,2}=\pm 1/2$). Hence the stretched H$_2$ provides the simplest physical realization of a molecular spin dimer. 

In order to excite spin dynamics in collinear spin-dimers, we have introduced a spatial inhomogeneity into the magnetic field pulses available in {\sc Octopus}. Thus, an inhomogeneous transverse magnetic field pulse of a few femptosecond duration is used to generate a spin misalignment in the dimer. In order to quantify such misalignment, we need a measure for the spin of overlapping atoms. Although the TDDFT spin-density distribution is well-defined at every instance, the spin state (and the charge) of an individual atom in a molecule or solid is not an observable. This of course prevents us to rigorously map the TDDFT dynamics onto a classical Heisenberg model. In fact, computing expectation values of local spin operators (e.g. $\braket{\hat{\vcr{S}}_1\cdot\hat{\vcr{S}}_2}$) from {\it ab initio} wave functions is not a trivial task~\cite{LocalSpin} and it has been recently shown that a continuum of valid local spin definitions exist~\cite{Cordoba}. In order to overcome such difficulty we have implemented an intuitive {\it rotating spin} approximation for decomposing the spatial spin-density distribution into atomic contributions, which is based on defining an appropriate linear transformation. The idea is to use the two extreme states, the HS and LS spin-density distributions, as reference points for decomposing the spin-density of any given non-collinear spin-state obtained through TDDFT evolution, assuming that it is simply a result of rigid rotation of some portion of the spin distribution in space. We will demonstrate that such method allows us to practically eliminate from the definition of the local spins the dependence on a particular spatial volume and that this can be done for a wide range of interatomic distances. This gives us the opportunity to define with a unique criterion the local spin trajectories, and thus to extract an effective exchange parameter $J$ for the spin dimer. Interestingly, the results agree quantitatively with those obtained by the broken symmetry method.  

All the TDDFT simulations are performed at the level of the adiabatic local spin-density approximation (ALSDA)~\cite{ALDA} with the modified parameterization of the correlation functional by Perdew and Zunger \cite{PZ}. The electron density and all the observables are represented over a dense real space grid (with a spacing of 0.1~\AA), and the entire simulation box is a parallelepiped of square cross-section (with a $\sim$12~\AA\ side) and a length (along the axis of the molecule) ranging between 20~\AA\ and 30~\AA, depending on the length of the molecule considered (in the dissociating limit). The time propagation of the Kohn-Sham equation is performed via the Crank-Nicholson (implicit mid-point) rule, while the Lanczos approximation of the propagator is used, as implemented in {\sc Octopus}~\cite{Castro04}. The typical time-step used in the simulations is 0.004~fs. 

We consider first a generic two-center magnetic molecular complex, a spin-dimer, as cartooned in Fig.~\ref{fig01}(a). The ground state DFT calculation is initialized in either the HS or the LS collinear configuration. In order to generate spin non-collinearity, a spatially-inhomogeneous external magnetic field pulse $\vcr{B}_\mathrm{ext}(\vcr{r},t)$ is applied to the TD-SDFT Hamiltonian \cite{Capelle}
\begin{eqnarray}
&\!\!H_{KS}\RnB{\vcr{r},t} = \\
&\!\!\sum_i^N \SqB{-\frac{\hbar^2\nabla_i^2}{2m}-\mu_B \vcr{\sigma}_i \cdot \vcr{B}_s(\vcr{r}_i,t)}\!\!\delta(\vcr{r}\!-\!\vcr{r}_i) + v_{s}\RnB{\vcr{r}}\,, \nonumber
\end{eqnarray}
where the sum runs over all ($N$) electrons in the system, $\vcr{B}_s=\vcr{B}_{xc}+\vcr{B}_\mathrm{ext}$ is the effective magnetic field for the KS orbitals, $\vcr{\sigma}_i$ is their spin operator, $\mu_B$ is the Bohr magneton and $v_s$ is the effective electrostatic potential which we assume not to carry an explicit time-dependence. We have implemented $\vcr{B}_\mathrm{ext}(\vcr{r},t)=\vcr{B}_0(\vcr{r}) \exp[-(t-t_0)^2/\tau_B^2]$ with a Gaussian time dependence and a variance $\tau_B$ typically between 2~fs and 5~fs. This is applied soon after the beginning of the time-dependent simulation ($t_0$ is chosen such that $\vcr{B}_\mathrm{ext}(t=0)$ is sufficiently close to 0 so that the discontinuity in the potential introduced at $t=0$ is negligibly small). For the spatial dependence $\vcr{B}_0(\vcr{r})$, we have experimented with a few simple continuous integrable functions and found that, as long as they are not symmetric with respect to the center of the molecule, there is little qualitative difference on the resulting spin dynamics. In other words, the sought outcome of spin-non-collinearity in the electronic structure is readily obtained for a wide range of $\vcr{B}_0(\vcr{r})$. In particular, we have found that there is no qualitative difference between a divergence-free solenoidal field, for instance, 
\begin{eqnarray}
&\!\!\vcr{B}_0(x,y,z) = B_0\mathrm{e}^{[-(x-x_0)^2/\xi^2]}\times\\
&\!\!\RnB{\vcr{e}_x+(x-x_0)y/\xi^2\vcr{e}_y + (x-x_0)z/\xi^2\vcr{e}_z}\nonumber
\end{eqnarray}
and the simplified $\vcr{B}_0(x,y,z)=B_0 \exp(-(x-x_0)^2/\xi^2)\vcr{e}_x$. Hence, in most of the simulations we have used the latter, where (typically) $x_0=-2$~\AA~with respect to the center of the molecule, $\xi=1$~\AA~and $\vcr{e}_x$ is a unit vector along the $x$-axis (aligned with the spin-dimer axis). This corresponds to a magnetic field transverse to the direction of the initial (ground-state) spin-polarization of the molecule, chosen as the $z$-axis [see Fig. \ref{fig01}(a)]. We have used values of $B_0$ ranging from 0.5~kT to 10~kT in order to generate desired misalignment for short enough simulation times. Clearly, these short, intense and very localized magnetic field pulses are only to be taken as theoretical tool for producing the misalignment, which onsets the spin dynamics.

Below we analyze the classical version of this problem, i.e. the dynamics of two misaligned classical angular momenta, $\vcr{S}_{1}$ and $\vcr{S}_{2}$, interacting according to Eq. (\ref{Jij}), and the possibility of mapping the TDDFT spin-density evolution onto that.

\myFig{1}{1}{true}{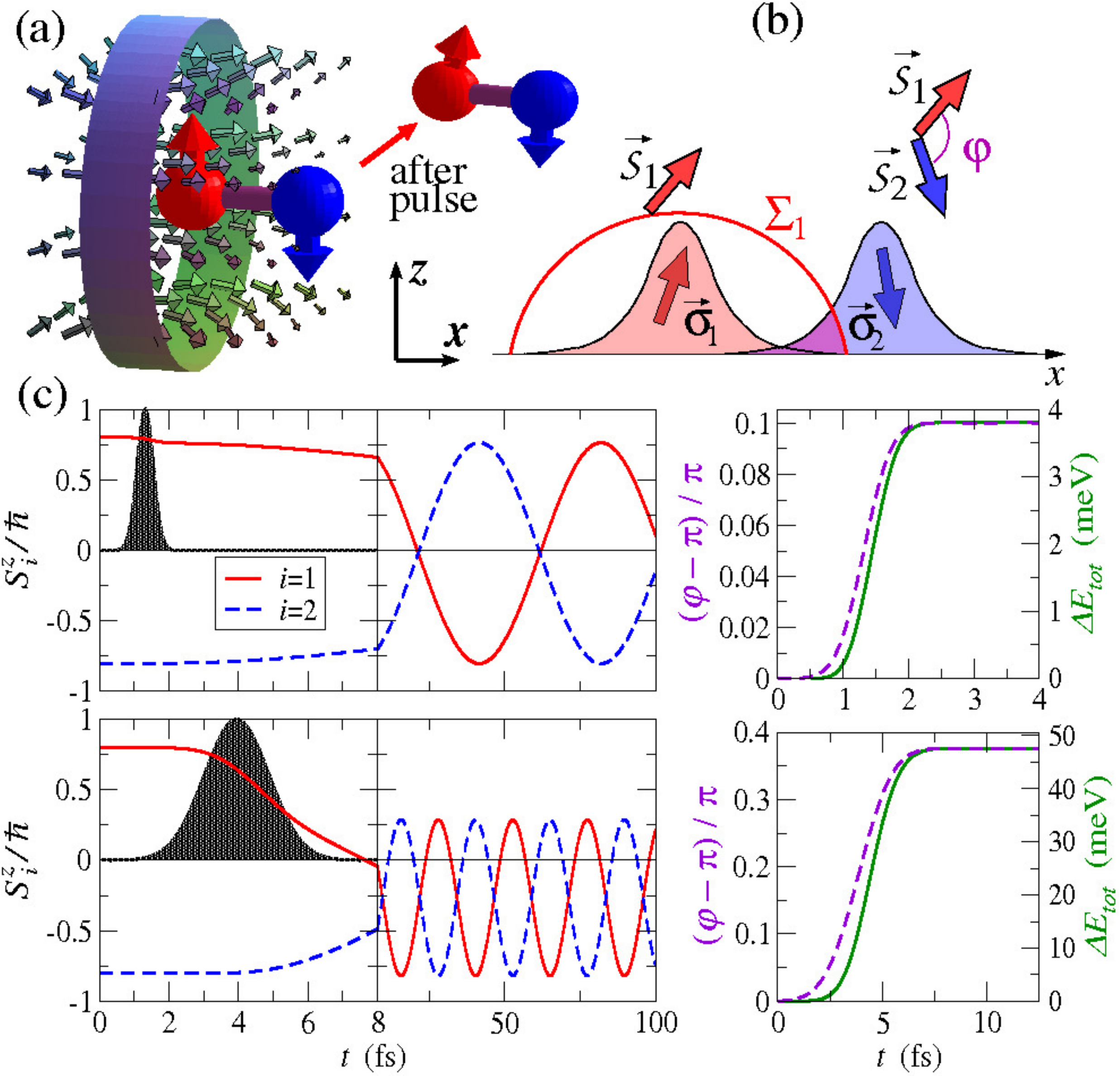}{(Color online) (a) Spin-dynamics is excited by a transverse magnetic field pulse (illustrated by the vector field). The ring represents a hypothetical solenoid with its center lying on the bond axis ($x$-axis) and offset from the mid point towards one of the atoms. (b) Illustration of the definition of local spin $\vcr{\cool{S}}_{1,2}$ [see Eq. (\ref{spin_sph_def})] and the definition of the angle $\varphi$. (c) Results from the {\it ab initio} time dependent simulations for H$_2$ and two different durations of the magnetic field pulse (shaded area): trajectories of the local spins' $z$-component $\cool{S}^z_{1,2}$, angle $\varphi$ between them and the expectation value of the total TDDFT energy of the system. The radius of the sphere defining $\vcr{\cool{S}}_{1,2}$ is $r_\mathrm{sph}=a/2$ and the bond length is $a=2.6$\,\AA.}{fig01}

\subsection{Classical Heisenberg model solution}

First, we examine the case of two rigid classical angular momenta $\vcr{S}_{1,2}$ interacting according to a Heisenberg spin Hamiltonian
\begin{equation} \label{H_cl}
H_{cl}=-2 J_\mathrm{cl}\vcr{S}_1 \cdot \vcr{S}_2\,. 
\end{equation}
Since we choose to have $\Abs{\vcr{S}_1}=\Abs{\vcr{S}_2}=S=1/2$, the factor 2 in equation~(\ref{H_cl}) is introduced in order for $H_\mathrm{cl}$ to produce the same difference between the energies of the parallel and antiparallel alignment of the classical spins as the triplet-singlet energy-difference $\Delta E_\mathrm{s-tr} = \langle \uparrow\uparrow | \hat{H} | \uparrow\uparrow \rangle - \langle \uparrow\downarrow | \hat{H} | \uparrow\downarrow \rangle = -J$ of the corresponding quantum spin Hamiltonian 
\begin{equation}
\hat{H}=-J \hat{\vcr{S}}_1 \cdot \hat{\vcr{S}}_2\:.
\end{equation}
In the classical spin Hamiltonian [Eq.(\ref{H_cl})] we include the physical dimension ($\hbar$) of the angular momenta in the coupling constant $J_{cl}$, which has a dimension of energy in analogy to the exchange parameter $J$. Hence, the classical equation of motion for each spin, say $\vcr{S}_1$, is
\begin{equation} \label{clEOM}
\dot{\vcr{S}}_1 = \CrB{\vcr{S}_1,H_\mathrm{cl}} = -2J_{cl} \sum_l S_2^l  \CrB{\vcr{S}_1,S_1^l} = \frac{2J_\mathrm{cl}}{\hbar} \vcr{S}_1 \times \vcr{S}_2 \,,
\end{equation}
where we have used the Poisson bracket for the corresponding classical angular momenta $\CrB{S^k\hbar,S^m\hbar}=\varepsilon_{klm}S^m\hbar$, with $\varepsilon_{klm}$ representing the fully anti-symmetric Levi-Civita tensor. For classical spins, forming an arbitrary angle $\varphi$, Eq.(\ref{clEOM}) describes a precessional motion about the total spin, $\vcr{S}_\mathrm{tot} \equiv \vcr{S}_1+\vcr{S}_2$, with an angular velocity
\begin{equation} \label{omega}
\omega = 4 J_\mathrm{cl} S \cos{\RnB{\varphi/2}}/ \hbar\,.
\end{equation}

Finally, we note that the precessional motion is stable against the application of any homogeneous external magnetic fields, which in the case of the quantum system can be used to define the quantization axis. In other words, if a homogeneous external magnetic field is applied, say along the $z$-axis, the total spin $\vcr{S}_\mathrm{tot}$ is driven into a precession about the field, but the individual classical spin components precess about the total spin with the same frequency given by equation (\ref{omega}). Hence, the trajectories of $S_1^{z}$ and $S_2^z$ are still harmonic oscillations at the field-free frequency $\omega$.

\subsection{Qualitative results for the {\it ab initio} spin-dynamics simulations}

The harmonic behavior, characteristic of the classical spin model described in the previous section, is easily obtainable in the {\it ab initio} spin-dynamics simulations of several spin dimers excited by an inhomogeneous magnetic field pulse. In fact, any spin-density component~\footnote{Note that this is valid for any component of the spin density, if there is no static homogeneous magnetic field applied. If there is, say a homogeneous field $\vcr{B}_h=(0,0,B_h)$, the result remains valid only for the spin-density component along the field, i.e. the $z$-axis, in this case.} integrated over any arbitrary finite volume in the simulation box shows a sinusoidal trajectory to a good accuracy for a number of periods [see, for instance, Fig. \ref{fig01}(c); our longest simulations have confirmed that observation for up to 10-12 periods; deviations from the ideal sinosoidal behavior in terms of higher frequency noise have been observed only in the case of small bond lengths and very small angles $\varphi$]. This seems to be the case for range of different two-center spin-polarized molecules, ranging from H$_2$ in a stretched (dissociating) configuration, to the hypothetical H-He-H trimer, and to much more electronically-complex high-spin entities like Mn$_2$ (not discussed here). 

In order to analyze in a quantitative way this numerical observation we consider first the most intuitive definition of local atomic spins: a local spin is obtained by integrating the spin density over non-overlapping spheres centered around each ion. In this way, from the instantaneous expectation value of the spin-density, a pair of Cartesian vectors, $\CrB{\vcr{\cool{S}}_1(t),\vcr{\cool{S}}_2(t)}$ can be extracted. As an example, the trajectories of the $z$-component of the spins obtained by integrating over spheres of radius half of the bond-length are presented in Fig. \ref{fig01}(c). We find these [e.g. $\cool{S}_1^z(t)$] to be sinosoidal after the extinction of the pulse ($t>\tau_\mathrm{pulse}=t_0+n\tau_B$, with typically $n=3$) and we are able to extract the angular velocity of precession $\omega_\mathrm{fit}$. Then, a characteristic {\it dynamical} exchange parameter can be evaluated from Eq.~(\ref{omega}) as
\begin{equation} \label{Jdyn}
J_\mathrm{dyn} = \frac{\omega_\mathrm{fit} \hbar}{ 4 \cool{S} \cos{(\varphi/2)}}\,,
\end{equation}
where $\varphi=\overline{\measuredangle \CrB{\vcr{\cool{S}}_1(t), \vcr{\cool{S}}_2(t)}}_{t>\tau_\mathrm{pulse}}$ is the angle between the local spins after the pulse and $\cool{S}= \overline{\Abs{\vcr{\cool{S}}_1(t)}}_{t>\tau_\mathrm{pulse}} = \overline{\Abs{\vcr{\cool{S}}_2(t)}}_{t>\tau_\mathrm{pulse}}$ is the averaged long-time local spin magnitude (which in all our simulations is practically identical between the two sites). These averaged quantities are typically very stable and independent on the length of the simulation. As is evident from the right-hand side panels of Fig. \ref{fig01}(c), after the decay of the pulse the angle $\varphi$ saturates to a constant (noise is typically in the fourth decimal place of the value in radians). 
 
\subsection{Defining the local spin} \label{secLinT}

Local (atomic) spins in DFT calculations are usually estimated through some sort of {\it partitioning} of the total density, for instance, the popular Mulliken and L\"owdin schemes. Typically, for calculations based on localized basis set, a population analysis consists in projecting over the chosen atomic orbital basis. The local spins are then extracted from the elements of the density matrix, contracted in spin space by the Pauli matrices. In the case of non-orthogonal bases these are weighed by the corresponding matrix elements of the square-rooted overlap matrix~\cite{Peralta}.   

In {\sc Octopus} a readily available implementation exists for evaluating local magnetic moments directly as integrals of the spin-density distribution,
$\vcr{\sigma} (\vcr{r})$,
\begin{equation} \label{spin_sph_def}
\vcr{\cool{S}}_i=\int_{\Sigma_i} \vcr{\sigma} (\vcr{r}) d\vcr{r}\:,
\end{equation}
over individual spherical volumes $\Sigma_i$ of radius $r_\mathrm{sph}$ centered around each atom $i$. We call this definition {\it direct} and the correspondent spins {\it apparent}. Because of the overlap of the atomic wave-functions associated to the individual atoms in the interstitial region, the value of the local spin at site 1, defined as Eq.~(\ref{spin_sph_def}), contains a contribution from site 2. This undesired contribution depends strongly on the radius $r_\mathrm{sph}$ [see Fig. \ref{fig01}(b)]. Hence, for instance, the {\it apparent}\, inter-spin angle $\varphi$ between two overlapping atoms is smaller than the actual angle between the overlapping atomic spin densities.

In order to decouple the contributions from the two sites, a simple linear transformation can be devised to eliminate the spatial dependence in the local spin definition. This is exactly true in the case of a uniform spin-density distribution of the individual overlapping sites. Let us assume that in the case of the dissociating hydrogen molecule, the $i$-th electron ($i=1,2$), predominantly localized on the site $\vcr{R}_i$, contributes to the total spin-density
\begin{equation} \label{sigma_r}
\vcr{\sigma}_i(\vcr{r})=f_{\sigma}(\vcr{r}-\vcr{R}_i)\vcr{\sigma}_i \,,
\end{equation}
where $f_{\sigma}(\vcr{r}-\vcr{R}_i)$ is integrable and confined to a compact (connected) spatial region. Note that this does not imply necessarily a minimal basis model where only the two 1$s$ atomic orbitals are considered. The function $f_\sigma(\vcr{r}-\vcr{R}_i)$ is the probability density distribution of the $i$-th electron, which does not need to be spherically symmetric. The vectors $\vcr{\sigma}_i$ are dimensionless and represent the {\it actual} spin direction (expectation value) of that electron, which in general is not directly observable from the DFT calculation. We can then chose a sphere $\Sigma_1$ that encloses most of that volume as (naively) cartooned in Fig. \ref{fig01}(b). Then, the apparent local-spins can be expressed as
\begin{eqnarray}
\vcr{\cool{S}}_1 &=&\alpha \vcr{\sigma}_1 + \beta \vcr{\sigma}_2 \nonumber \\
\vcr{\cool{S}}_2 &=&\beta \vcr{\sigma}_1 + \alpha \vcr{\sigma}_2 \, ,
\end{eqnarray}
where $(\alpha, \beta)=\int_{\Sigma_1}f_{\sigma}(\vcr{r} \pm \vcr{d}/2)d\vcr{r}=\int_{\Sigma_2}f_{\sigma}(\vcr{r} \mp \vcr{d}/2)d\vcr{r}$ (the top signs are for $\alpha$ and the bottom ones for $\beta$) and $\vcr{d}\equiv d\,\hat{\boldsymbol{x}}$ is the bond length, which is aligned along the $x$ axis for definiteness. In other words, $\vcr{\sigma}_i$ can be determined from the inverse of the above linear transformation as
\begin{equation} \label{lintransf}
\RnB{\begin{array}{c} \sigma_1^l \\ \sigma_2^l \end{array}}  = A \RnB{\begin{array}{c} \cool{S}_1^l \\ \cool{S}_2^l \end{array}} \, ,
\end{equation}
where 
\begin{equation}
A \equiv \RnB{\begin{array}{cc} \alpha & \beta \\ \beta & \alpha \end{array} }^{-1} =  \RnB{\begin{array}{cc} a & b \\ b & a \end{array} }
\end{equation}
for any Cartesian component $l\in \CrB{x,y,z}$. As $\alpha$ and $\beta$ are in principle unknown, $A$ can be determined from the calculated apparent spins in the collinear configurations. Let $\cool{S}_{\updo}$ and $\cool{S}_{\upup}$ be the apparent local spin values in the singlet and in the triplet (broken symmetry) state, respectively, and we consider normalized $\vcr{\sigma}_i$, i.e. $A$ needs to fulfill the following equation
\begin{equation}
\cool{S}_{\updo} A \RnB{\!\!\begin{array}{r} 1 \\ -1 \end{array}\!\!} = \RnB{\!\!\begin{array}{r} 1 \\ -1 \end{array}\!\!} \quad \mathrm{and} \quad  \cool{S}_{\upup} A \RnB{\!\begin{array}{r} 1 \\ 1 \end{array}\!} = \RnB{\!\begin{array}{r} 1 \\ 1 \end{array}\!} \,.
\end{equation}
The matrix elements of $A$ that satisfy this requirement are
\begin{equation}
a=\frac{\cool{S}_{\updo}+\cool{S}_{\upup}}{2\cool{S}_{\updo}\cool{S}_{\upup}}, \qquad b=\frac{\cool{S}_{\updo}-\cool{S}_{\upup}}{2\cool{S}_{\updo}\cool{S}_{\upup}} \,.
\end{equation}

Hence, through $A$ the individual electronic (and atomic in the case of hydrogen) spin polarization directions $\vcr{\sigma}_{1,2}$ can be worked out from the apparent (sphere-integrated) local spin quantities $\vcr{\cool{S}}_{1,2}$. The practical applicability of this definition to the dynamically generated non-collinear spin configurations depends on how small the actual redistribution of electron charge between the HS and LS collinear states is. That is, how close the individual electron charge distribution $f_{\sigma}(\vcr{r})$ is to a constant of motion for the particular {\it ab initio} spin-dynamics simulation. In other words, if the dynamics can be locally described by an inter-rotation of overlapping spin-density kernels without a local norm variation, the spatial factor in the definition of the local spins can be completely eliminated. This might also be considered as an approximation, providing grounds for an alternative density-based definition of local spin expectation values, which significantly reduces the effects of overlap inherent to the directly space-integrated atomic quantities. We will call $\vcr{\sigma}_{1,2}$ the {\it transformed} local spins. It will be demonstrated in the following sections that, with regard to the Heisenberg interaction, this constitutes a good approximation for the simplest spin-dimer systems up to considerably small bond-lengths where the atomic overlap is significant.

\section{Implementation and calculations for H$_2$}

The typical outcome of the described above TDSDFT simulations of a stretched H$_2$ molecule (at a bond-length $d=2.6$ \AA) is presented in Fig. \ref{fig02} in the form of 2-dimensional contour plots representing the stacked snapshots of the various observables (expectation value) distribution along the molecule axis as a function of the simulation time (in the horizontal direction). This visualization offers a quick glimpse of the dynamics. For instance, it shows that the inhomogeneous magnetic field pulse, used to generate non-collinearity from the LS ground state, produces a localized spin and charge redistribution. A comparison between the pulse geometry in  Fig.~\ref{fig02}(a) and the charge and spin-currents in the underlying graphs shows little direct spatial correlation (e.g. the pulse is centered at -2\,\AA\, while the excitation is centered at -1.3\,\AA\ where the proton sits) and this demonstrates further the freedom available in the choice of the actual magnetic field distribution~\footnote{We find that as long as the magnetic field is not symmetric with respect to the center of the dimer this qualitative result persists.}. It also shows that the small relative charge and spin redistribution in the dimer follows closely the temporal shape of the pulse. After the the pulse dies out, only a tiny amount of charge sloshing between the two sites at very high frequency remains, as evident from Fig. \ref{fig02}(b). The figure represents the charge current as the sum of the up-spin and down-spin components (with respect to the quantization axis set by the initial spin-polarization at $t=0$) of the expectation value of spin current tensor in the direction of the bond ($x$-axis). The later currents are defined by only two scalar components $J^\uparrow_x \equiv \vcr{J}^{\uparrow\uparrow}_{x}$ and $J^\uparrow_x \equiv \vcr{J}^{\uparrow\uparrow}_{x}$ of the spin-current tensor 
\begin{equation}\label{spintensor}
\vcr{J}^{\alpha\beta}_{l} (\vcr{r}) =  \sum_n \braket{\vcr{\sigma}_n^{\alpha\beta}\otimes j_n^l(\vcr{r})} \,,
\end{equation}
where $l\in\CrB{x,y,z}$, $\hat{\vcr{j}}_n(\vcr{r})=\frac{\hbar}{2mi}\RnB{\vcr{\nabla}_n\delta(\vcr{r}-\vcr{r}_n)-c.\,c.}$ is the orbital current operator for the $n$-th electron and we have omitted the implicit time-dependence for simplicity. Note that while the charge current is near to zero after the pulse, the spin-current builds up. After the pulse-coherent depletion of the longitudinal spin in the site more exposed to the pulse (the site at -1.3\,\AA), a uni-directional spin-current is established. This corresponds to a transfer of spin-up along the positive $x$-direction and of spin-down along the negative $x$-direction. Hence, the the up-spin localized at the left site starts turning down while the down-spin on the right starts turning up. Figure \ref{fig02}(d,e) shows that while this spin-rotation process is taking place the distribution of the both the charge and the magnitude of the spin-density after the pulse tend to remain stationary in space.

\myFig{1}{1}{true}{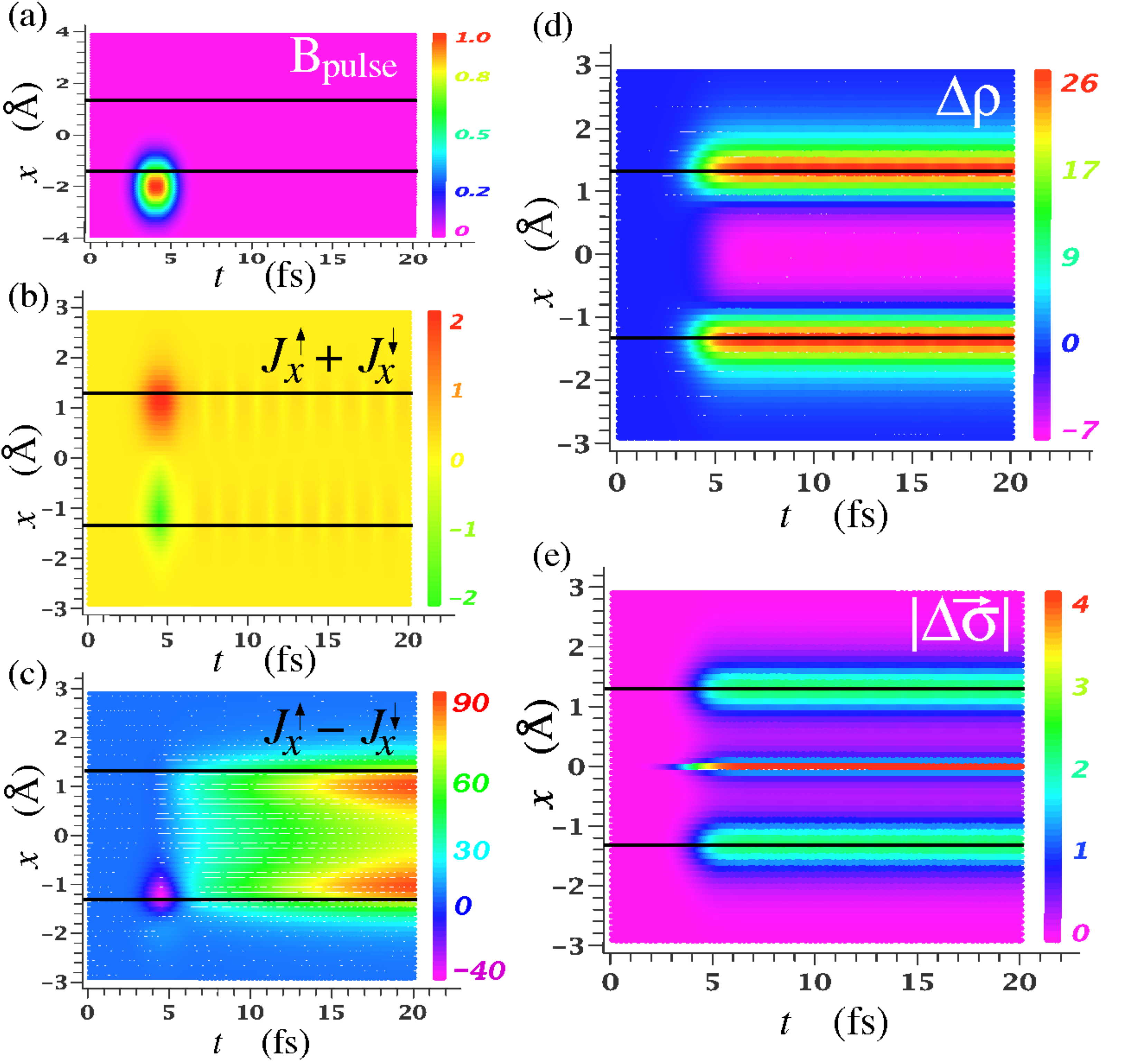}{(Color online) Contour plots of the time evolution (early-time) of the distribution along the direction of the bond ($x$) of (a) the external magnetic pulse, in units of 3\,kT; (b) the charge current density and (c) the $z$-component of the spin-current tensor [see Eq. (\ref{spintensor})] in the same arbitrary unit scale; and the variations with respect to the ground state of (e) the charge density $\Delta \rho(x,t)\equiv\rho(x,t)-\rho(x,0)$, in units of $0.03e/\mathrm{\AA}^3$; and (d) the magnitude of the spin density (Euclidean norm) $\Abs{\Delta \vcr{\sigma}(x,t)}=\Abs{ \vcr{\sigma}(x,t)} - \Abs{\vcr{\sigma}(x,0)}$, in units of $0.3(\hbar/2)/\mathrm{\AA}^3$. Note that soon after the magnetic pulse dies out the system becomes nearly stationary with respect to charge transfer between the two sites.}{fig02}

In the long time limit, as a result of the such generated non-collinearity, a regular pattern of rigid local spin rotation is established throughout space (spin-density at every point in the simulation box precesses about the total spin of the dimer with the same frequency). This is also evident from the sloshing of pure spin currents between the two atoms (see Fig.~\ref{fig03}). The corresponding trajectories of the spin-density integrated over atomically-centered spheres are similar to those depicted in Fig.~\ref{fig01} (for a different pulse strength) and are typically sinusoidal to a great level of accuracy within the duration of simulation (up to 200-250\,fs).

\myFig{1}{1}{true}{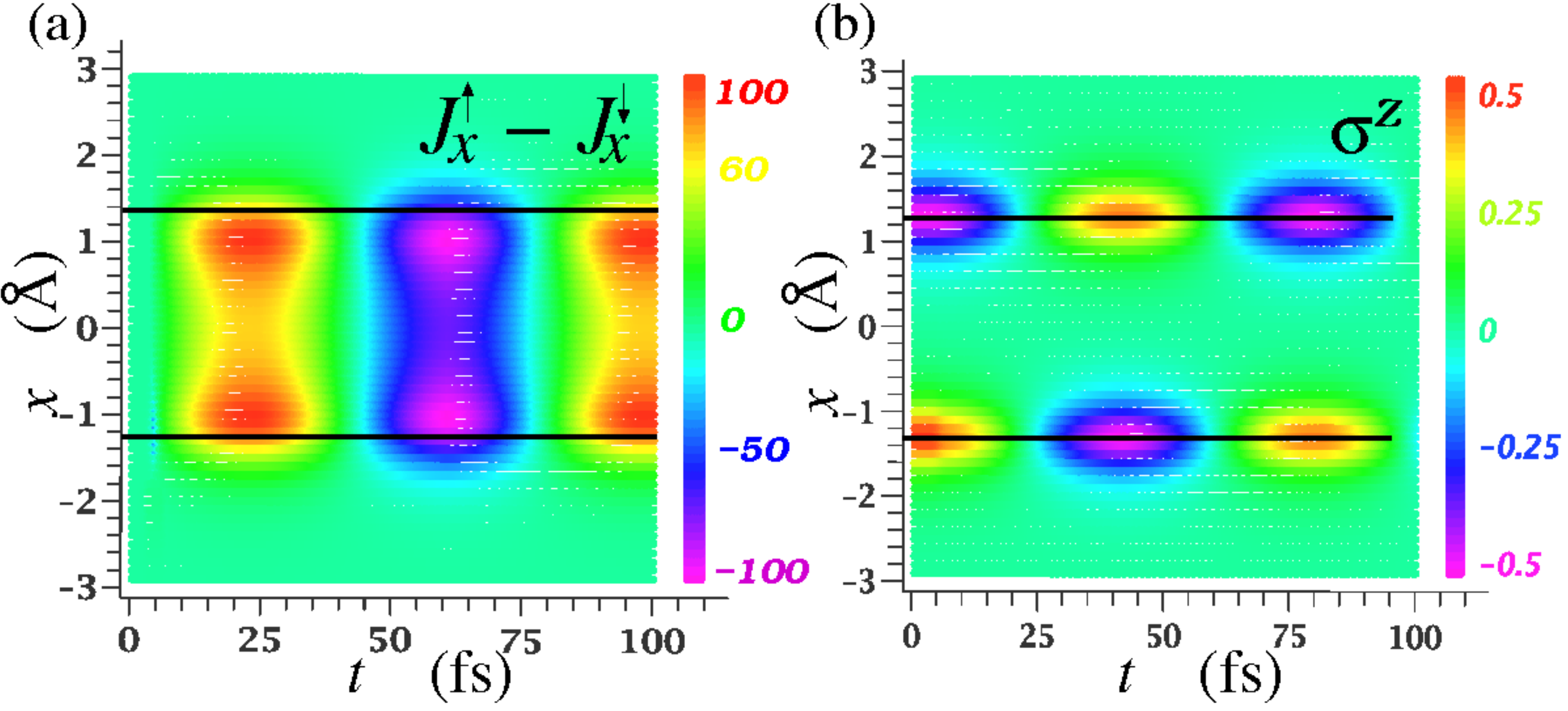}{(Color online) Contour plots of the time evolution (long-time limit) of the distribution along the direction of the bond ($x$) of (a) the $z$-component of the spin current density tensor (arbitrary units, as in Fig.\ref{fig01}) and (b) spin density $z$-component, in units of $300(\hbar/2)/\mathrm{\AA}^3$.}{fig03}

The properties of the linear transformation $A$ and with respect to the mapping of the H$_2$ spin-dimer dynamics onto classical degrees of freedom are demonstrated in Fig.~\ref{fig04}. A set of non-collinear quasi-stationary dynamical states with angles $\varphi_\mathrm{sph}$ between the {\it apparent} local spins ranging from 0 to $\pi$ are produced by applying pulses of different strengths from either the LS or the HS initial state. We find a systematic variation of the {\it apparent} local spin norm, $\cool{S}$, as a function of $\varphi_\mathrm{sph}$ alone, irrespective of the particular collinear initial state. The effect of the sphere radius on that dependence is significant [see Fig.~\ref{fig04}(a)]. In contrast, for the {\it transformed} spins the quality of the achieved normalization of $\vcr{\sigma}_{i}$ is quite high regardless of the angle $\varphi$ (even when this is close to $\pi/2$), and is practically the same for any size of the integration sphere [Fig.~\ref{fig04}(c)]. The magnitude of the {\it transformed} local spin varies by at most 0.5\%. Such variation is practically negligible on the background of the variation of the apparent spin magnitude $\cool{S}$ at the two extreme HS and LS states as a function of $r_\mathrm{sph}$ [see Fig.~\ref{fig04}(d)]. At the same time the transformation of the angle shows a small but systematic dependence on the sphere radius [see Fig.~\ref{fig04}(b)]. The larger the spheres, the more of the overlap they capture and the greater the correction in the angle $\varphi$ achieved by the transformation.

\myFig{1}{0.8}{true}{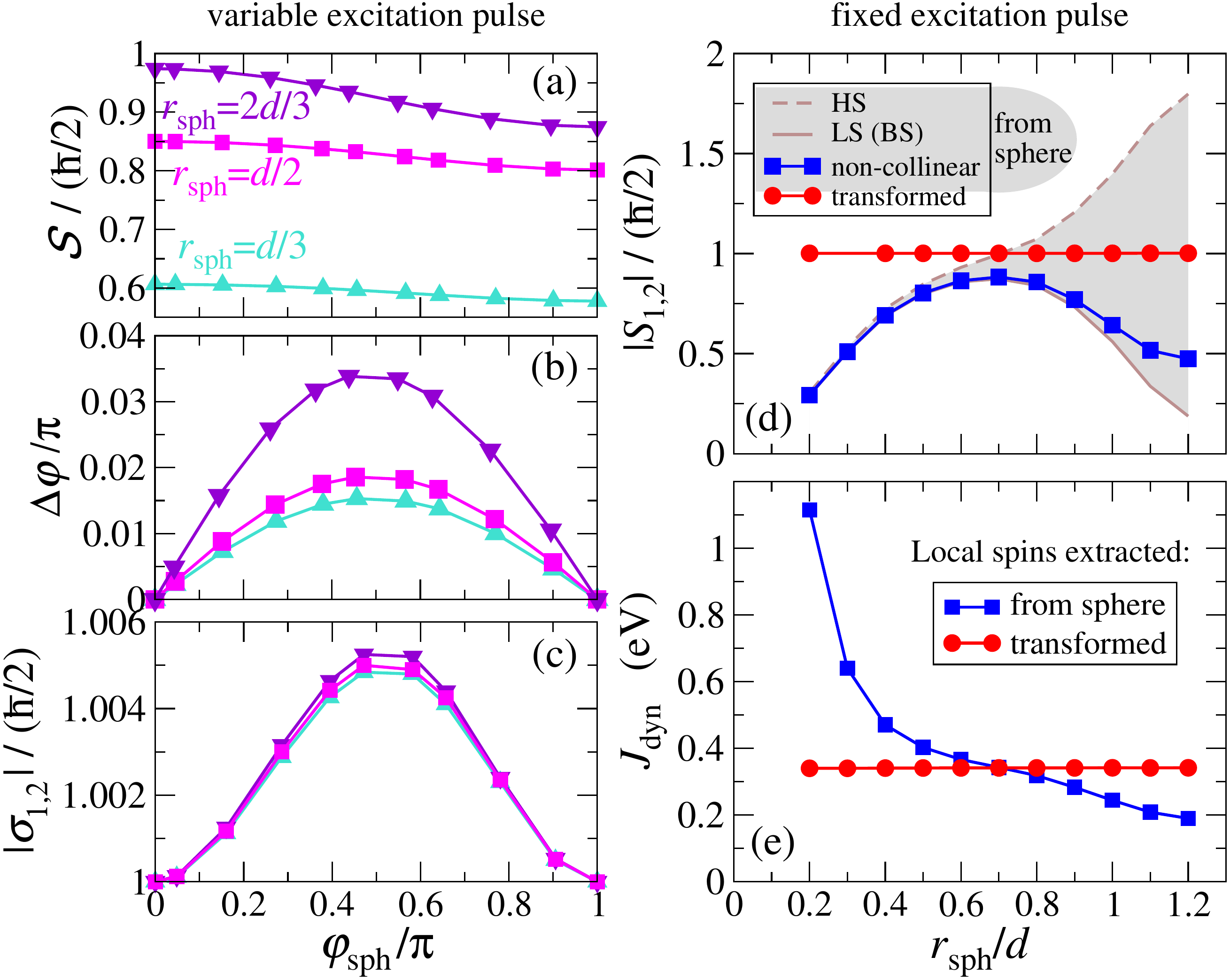}{(Color online) Left hand side panels: constants of motion (numerically acceptable) as a function of the apparent spin misalignment angle $\varphi$ after the pulse for three different radii of the sphere (a) the magnitude of the apparent local spin $\cool{S}$; (b) the difference between transformed and direct angle between local spins $\Delta \varphi  = \varphi - \varphi_\mathrm{sph}$; (c) transformed local spin magnitude $\Abs{\vcr{\sigma}_{1,2}}$. Right hand side panels: as a function of the sphere radius (d) local spin magnitude in the two stationary collinear states and in the non-collinear state, defined over the sphere ($\cool{S}$) and the transformed non-collinear spin magnitude ($\Abs{\vcr{\sigma}_{1,2}}$) for the largest sphere; (e) corresponding Heisenberg constants $J_\mathrm{dyn}$, as defined by Eq. (\ref{Jdyn}). These caclulations are for H$_2$ molecule of bond length $d=2.6$\,\AA.}{fig04}

The calculated angle between the transformed local spins, $\varphi$, for any value of $r_\mathrm{sph}$, is always above the upper asymptotic limit of the apparent $\varphi_\mathrm{sph}$ as function of the sphere radius. We find that $\varphi_\mathrm{sph}$ always tend to a saturation maximum for decreasing $r_\mathrm{sph}$. In fact, for the particular excitation depicted in the right-hand-side panels of Fig. \ref{fig04}, $\varphi$ varies just between 2.648\,rad and 2.650\,rad for $r_\mathrm{sph}$ ranging between $0.2d$ and $1.2d$, while the change in the apparent angle $\varphi_\mathrm{sph}$ is massive, i.e. it changes from 2.626\,rad down to 0.789\,rad (this data is not presented on the graph). Figure \ref{fig04}(e) shows the resulting correction in the corresponding exchange parameter $J_\mathrm{dyn}$, defined as in Eq. (\ref{Jdyn}). It demonstrates that using the {\it apparent} local spins for $J_\mathrm{dyn}$ is completely meaningless: the dependence on the sphere radius is very strong (for instance, in the large radius limit $J_\mathrm{dyn}$ understandably tends to 0). In contrast by using the transformed quantities, $J_\mathrm{dyn}$ as a function of $r_\mathrm{sph}$ is constant with an accuracy of less than 0.15\%  (for the case depicted in Fig. \ref{fig04}, $J_\mathrm{dyn}=0.3413\pm 0.0005$\,eV averaged for the 11 values of $r_\mathrm{sph}$ in the range from $0.2d$ to $1.2d$). Note that the linear transformation does not change the observed angular frequency of local spin rotation $\omega_\mathrm{fit}$. This is because any spatial portion of spin density in the non-collinear state rotates at the same rate.

\myFig{1}{1}{true}{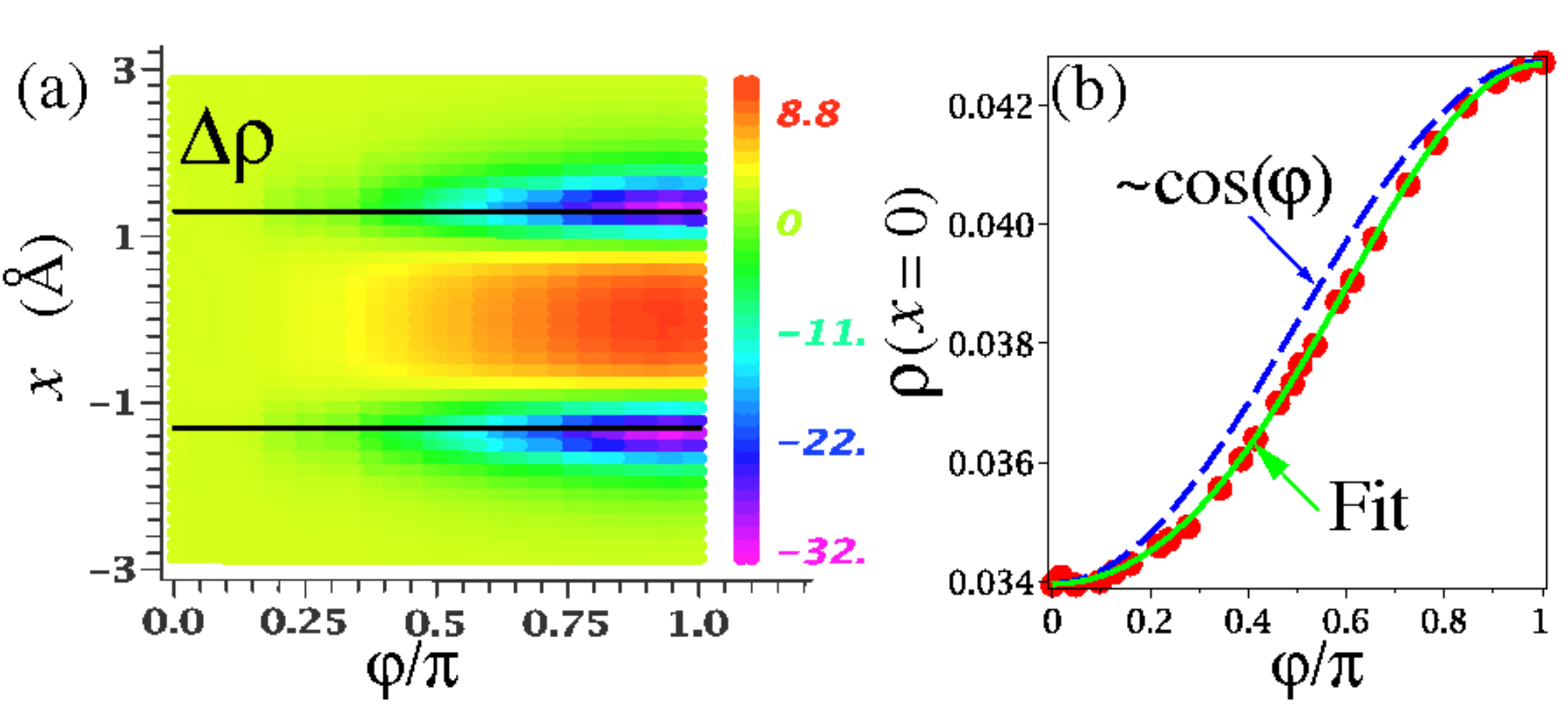}{(Color online) Dependence on the dynamically-generated angle $\varphi$ of (a) the variation of the electron density distribution along the bond axis $x$ with respect to the low-spin state, here $\Delta \rho (\varphi,x) = \rho (\varphi,x) - \rho (0,x) $, locations of the nuclei are marked by the black lines; (b) the electron density at $x=0$, compared to a cosine function (blue dashed curve) and fitted (least-squares) by a two-parameter function: $A\cos \varphi + B\sin^2\varphi$ (green curve). The units for $\rho$ and $\Delta\rho$ are $0.3e/\mathrm{\AA}^3$.}{fig05}

In order to gain more insight into the dynamically-achieved quasi-stationary non-collinear state of the spin dimer, we can look at the snapshots of the long-time limit electron density distribution as we systematically increase the strength of the excitation. In Fig. \ref{fig05}(a) the long-time charge density along the bond axis is presented as a function of the (transformed) inter-spin angle $\varphi$ (relative to the density of the HS state with $\varphi=0$). We find a clear visual evidence for the action of Pauli exclusion principle and the corresponding exchange-correlation hole. The contour plot shows that the HS state ($\varphi=0$) bond is depleted with respect to the LS state ($\varphi=\pi$). The dependence of the averaged charge density in middle of the dimer as a function of $\varphi$ is shown in Fig. \ref{fig05}(b). This nearly fits to a cosine function but not exactly. In fact, by including even only a second order harmonic ($\propto \cos2\varphi)$ from the Fourier series or a term proportional to $\sin^2\varphi$ (the two are the same up to an additive constant) a significant improvement of the fit is obtained.  

\myFig{1}{1}{true}{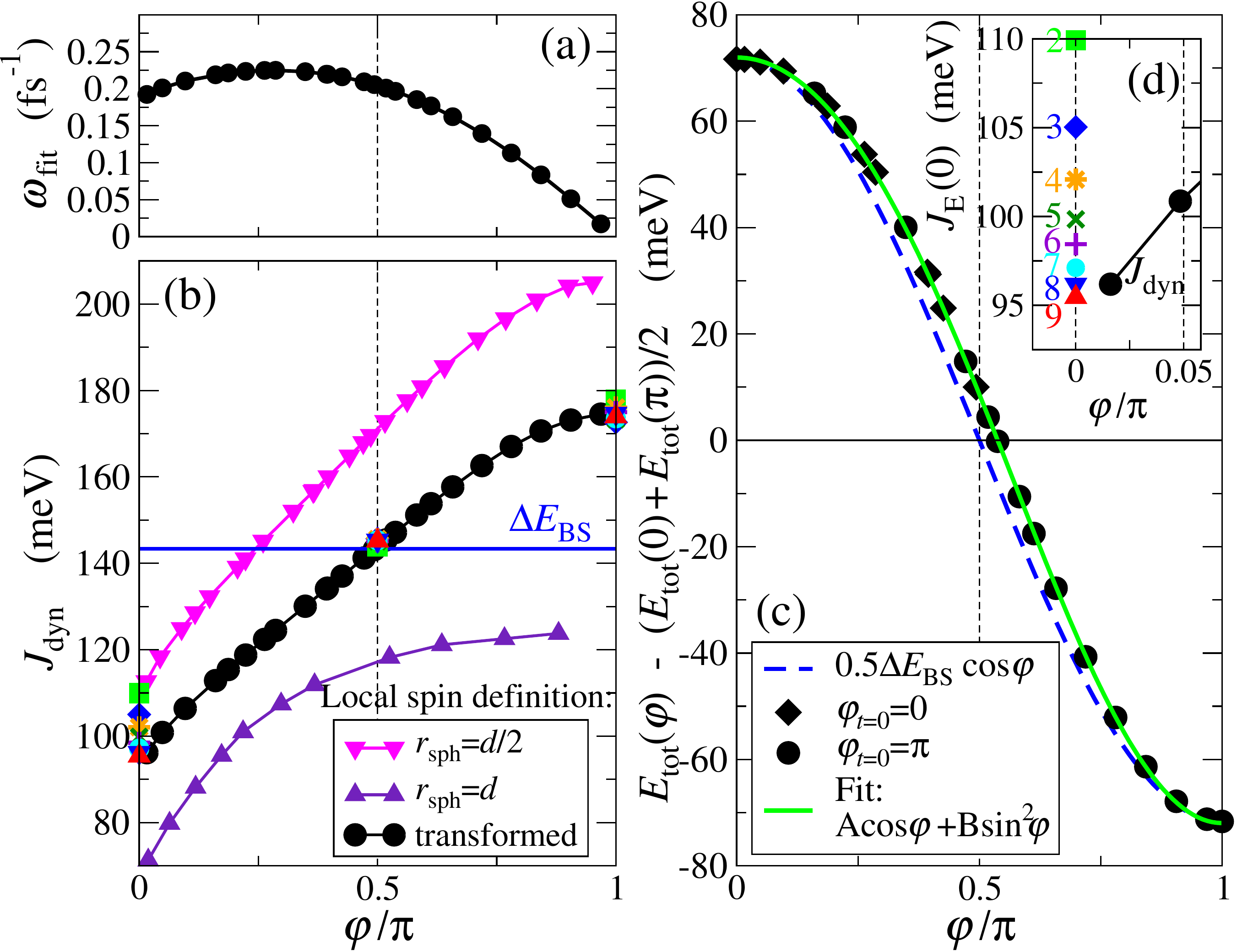}{(Color online) Dependence on the dynamically-generated angle, $\varphi$, of (a) the extracted frequency of rotation, $\omega_\mathrm{fit}$, (this is the same for the {\it apparent} as well as the {\it transformed} spins); (b) $J_\mathrm{dyn}$ extracted from the spin trajectories [Eq. (\ref{Jdyn})]; (c) the total TDDFT energy (long-time value) with respect to the average $E_\mathrm{tot}$ value between the LS and the HS states, i.e. $E_\mathrm{tot}(\varphi)-(E_\mathrm{HS}+E_\mathrm{LS})/2$; (d) consecutive approximations to $J_\mathrm{dyn}(\varphi=0)$ from Eq (\ref{JE}) using the series $\sum_{i=1}^n c_i \mathrm{cos}^i(\varphi)+c_0$ of increasing order $n=2,9$ as a fitting function for $E_\mathrm{tot}(\varphi)$. Marked in panel (b) is also $\Delta E_\mathrm{BS}=E_\mathrm{HS}-E_\mathrm{LS}=143.4$\,meV and the results for the exchange constant based on $E_\mathrm{tot}(\varphi)$ derivatives as in Eq~(\ref{JE}) (colored symbols at $\varphi=0$ and $\varphi=\pi$). The results presented here are for a stretched H$_2$ molecule ($d=2.6$~\AA).}{fig06}

We now move to analyze how the exchange parameter is calculated from the dynamical simulations initiated with different magnetic pulses, i.e. for different angles, $\varphi$. In Fig.~\ref{fig06}(a) we present the precession frequency $\omega_\mathrm{fit}$, while in Fig.~\ref{fig06}(b) the dynamical exchange parameter, $J_\mathrm{dym}$, as calculated from Eq.~(\ref{Jdyn}). Clearly both $\omega_\mathrm{fit}$ and $J_\mathrm{dyn}$ depend sensitively on the angle between the two spin moments. This is expected for $\omega_\mathrm{fit}$, which presents the same asymptotic behavior of the classical Heisenberg model for $\varphi\rightarrow\pi$. However $J_\mathrm{dyn}$ is not constant with $\varphi$, indicating that our quantum system, simulated with TDDFT, deviates from the classical one. Intriguingly the dynamical exchange seems to agree perfectly with that calculated from the broken symmetry approach, $\Delta E_\mathrm{BS}=E_\mathrm{HS}-E_\mathrm{LS}$, for $\varphi=\pi/2$, i.e. when the two local spins are orthogonal to each other. Note also that, once again, the {\it apparent} local spins cannot be used here since the variations in $J_\mathrm{dyn}$ with the choice of integration radius are very large.

A similar deviation from the classical Heisenberg model is found in the dependence of the total energy, $E_\mathrm{tot}$, on the angle $\varphi$ between the {\it transformed} local spins. Note that here we consider $E_\mathrm{tot}$ in the long-time limit, i.e. long after the external field pulse has extinguished. In this limit $E_\mathrm{tot}$ is a constant of motion with numerical fluctuations at times of about 100\,fs being typically smaller than 10$^{-6}$\,eV. The dependence of the total energy $E_\mathrm{tot}(\varphi)$ clearly deviates from the characteristic cosine form of the classical Heisenberg model [Fig. \ref{fig06}(c)]. However, as we have found for the charge density [see Fig.~(\ref{fig05})], also for $E_\mathrm{tot}$ the best fit of the dynamical quantities is obtained by including higher harmonics, with already a remarkably good agreement at the level of the second harmonic ($\propto\sin^2\varphi$, note that with the use of $\sin^2\varphi$ the offset of the fit is 0). The deviation of the total energy from the Heisenberg model ($\propto \cos\varphi$) can be attributed to a combination of factors. The Heisenberg model returns the energy of two localized electrons as the scalar product of their corresponding local spin operators~\cite{Yosida}. Clearly, any definition of the local spins in terms of the expectation values of spin-density and the corresponding $\varphi$ is an approximation. In addition, the total energy of the system in the non-collinear spin state is approximated by the choice of the LSDA for the exchange-correlation potential. 

A similar deviation from the Heisenberg model was found also for a few spin-dimer complexes by Peralta and Barone~\cite{Peralta}, who noticed that a systematic improvement of the agreement between the DFT results and the classical Heisenberg model is achieved as the approximation for the exchange and correlation functional improves. In particular a more Heisenberg-like behavior is found for hybrid functionals, such as B3LYP~\cite{B3LYP}. This is somehow expected, since in hybrid functionals the spurious self-interaction, which is present in LSDA, is partially removed and the electron charge gets more localized at the nuclear sites~\cite{Akin}. In brief, hybrid functionals return an electronic structure closer to that underpinning the classical Heisenberg model. In any case, a variation of $J$ (evaluated from the second derivative of the total energy with respect to $\varphi$) between the values calculated around the LS state or those around the HS one was found for all functionals. This variation has the same direction as our corresponding quantity, calculated as
\begin{equation} \label{JE}
J_\mathrm{E}(\varphi) \equiv \frac{1}{2S^2} \frac{d^2 E_\mathrm{tot} (\varphi)}{d\varphi^2} \cos (\varphi) \,.
\end{equation}

From figure~\ref{fig06}(c) it appears that $\Abs{J_\mathrm{E}(0)} > \Abs{J_\mathrm{E}(\pi)}$, since the total energy as a function of $\varphi$ is above the Heisenberg $\cos$-type dependence both at $\varphi=0$ and $\varphi=\pi$. We now demonstrate that this variation is consistent quantitatively with the exchange couplings $J_\mathrm{dyn}$ extracted from the dynamical trajectories via Eq.~\ref{Jdyn}. In fact, if we increase of the number of harmonics in the Fourier series fitting $E_\mathrm{tot} (\varphi)$, the match between the values of $J_\mathrm{E}$ and $J_\mathrm{dyn}$ calculated at the LS and HS states improves systematically [see Fig.~\ref{fig06}(d), inset of Fig.~\ref{fig06}(c)]. In other words, the spin dynamics of the molecule, excited and mapped out as described, indeed probes the landscape of the spin-dependent energy of the system and in the vicinity of the HS and LS state the agreement of $J_\mathrm{dyn}$ and the Heisenberg model approximation of the total energy is remarkable. In conclusion we find that the Heisenberg spin interaction is the governing mechanism for the {\it ab initio} spin dynamics of stretched H$_2$, although with some deviations. In particular $E_\mathrm{tot} (\varphi)$ contains higher contributions in the harmonic series over $\varphi$, beside the Heisenberg-type $\cos\varphi$ dependence.

\subsection*{Variation with distance}

The hydrogen molecule has had a special role in quantum chemistry as a basis for understanding the chemical bond. It is well known that the Heitler-London theory of molecular bonding incorrectly produces a spin-triplet ground state in the dissociation limit~\cite{Yosida}, because it omits the electron correlations. In the other limit, the Hartree Fock molecular orbital wave-function fails due to an overestimation of the ionic contribution. The ground state of the dissociating H$_2$ has a significant multi-configurational character and it is still an unsolved problem for DFT \cite{H2DFT}. Furthermore, the problem for the exchange coupling in H$_2$ is the one for the spin-flip excitation energy $^1\Sigma^+_g \rightarrow ^3\Sigma^+_u$. The standard ALDA in TDDFT is found to have severe weaknesses and it badly underestimates the excitation energies in the dissociation limit \cite{H2TDDFT1,H2TDDFT2}. Limiting ourselves to the non-collinear ALDA, the aim of our work is not to offer an accurate alternative evaluation of the exchange coupling in H$_2$, but to demonstrate a first attempt to relate the Heisenberg $J$ to the actual spin trajectories calculated from TDDFT. It is well-known that LSDA has serious shortcomings in describing long-distance exchange and correlation effects~\cite{H2DFT} and our dynamical analysis cannot improve on these. For the sake of completeness in Fig.~\ref{fig07} we present our results for the distance dependence of the exchange coupling in H$_2$ at medium distances (2-3 \AA) and compare it with a number of previously published calculations, obtained at various levels of approximation. 

\myFig{1}{1}{true}{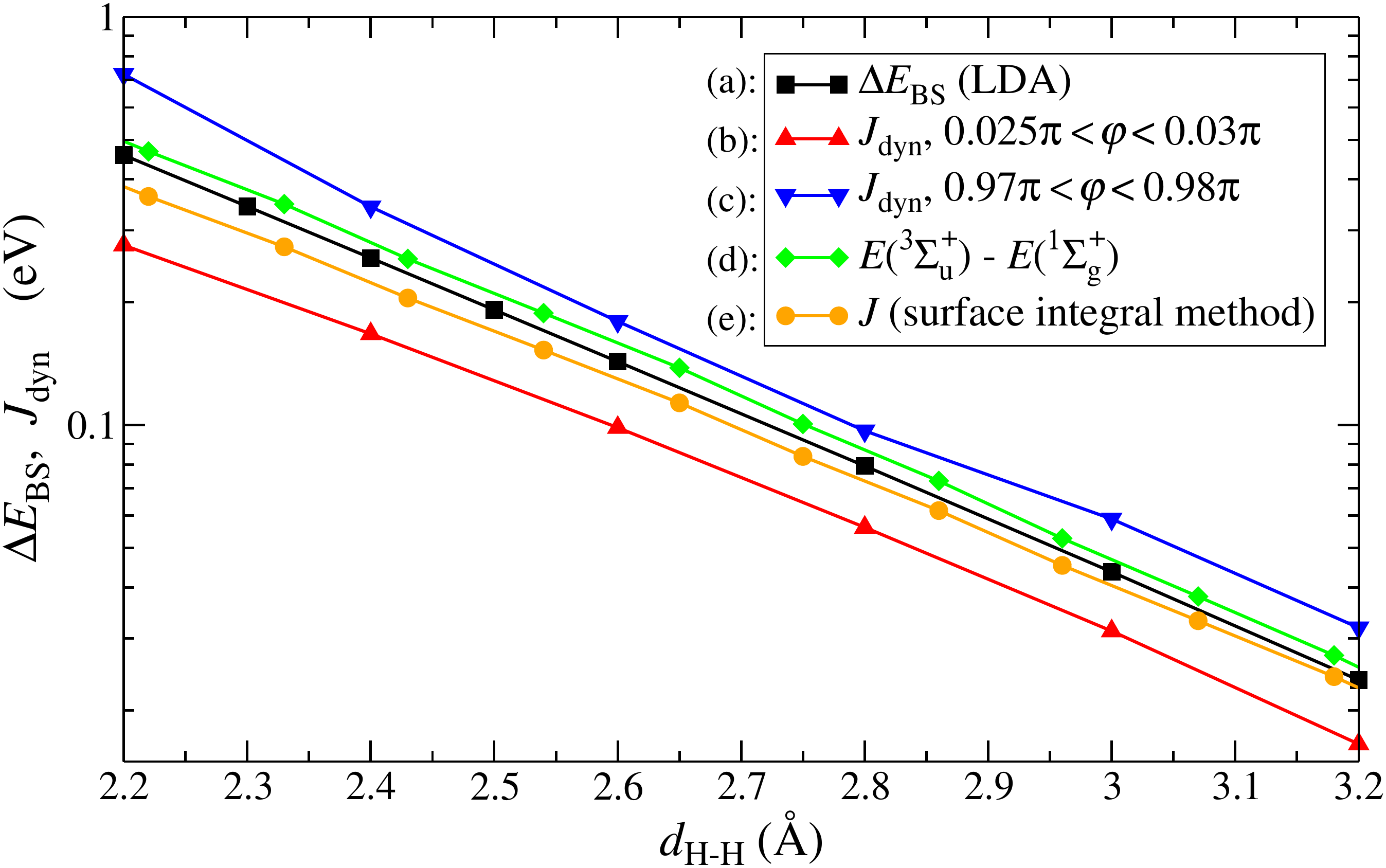}{(Color online) Comparison of the distance dependence of the exchange coupling in H$_2$ calculated as (a) broken symmetry energy difference from static LSDA; (b and c) $J_\mathrm{dyn}$ from Eq. (\ref{Jdyn}) for very small and very large angles $\varphi$, respectively; (d) {\it ab initio} variational calculation of ground state and first excited state total energies by Kolos and Wolniewicz \cite{Kolos}; (e) the leading term in the surface integral method by Herring and Flicker \cite{Herring}.}{fig07}

Our static broken symmetry LDA result lies nearly in the middle between the leading term in the perturbative calculation of $J$ obtained with the surface integral method~\cite{Herring} and the exact variational result for the first excitation energy~\cite{Kolos}. As discussed above, the value of our dynamical Heisenberg parameter, $J_\mathrm{dyn}$, depends strongly on the angle $\varphi$. We show the range of $J_\mathrm{dyn}$ values between some of the smallest and some of the largest angles obtained (pulses are purposely chosen as to produce angles of nearly the same magnitude for all $d$). The range is significant and it is relatively similar for all the bond-lengths. Notably, broken symmetry LSDA value at any distance is always well reproduced by our dynamical calculation for angles $\varphi \approx \pi/2$.  

\section{Results for the H-H\lowercase{e}-H trimer}

We now apply the dynamical scheme discussed so far to another system, namely the hypothetical H-He-H molecule. This is the simplest possible model system presenting a high order spin exchange interaction, e.g. the two H atoms interact via superexchange mechanism~\cite{Yosida} across the close shell He atom. There are no experimental observations for H-He-H but it is a good test case for new quantum chemistry methods~\cite{Ruiz} as full configuration interaction calculations exist \cite{Hart,Bencini} for comparison. Here, as many other works in the literature, we consider as typical the H-He distance of 1.625~\AA.

In general our results for H-He-H are similar to those for H$_2$. Again, after the application of the spatially-asymmetric magnetic field pulse the spins of the hydrogen electrons become misaligned by an angle $\varphi$ and start to precess about the total spin at a steady angular frequency. Fig.~\ref{fig08} shows the long-time oscillations of the spin-density and the oscillating $z$-polarized spin-currents along the bond axis. The spin current distribution is qualitative different from that of H$_2$ in Fig.~\ref{fig03}, since it now peaks at the He atom instead of the sites bearing the localized spins. This provides an insight of the indirect exchange mechanism in H-He-H. 

\myFig{1}{1}{true}{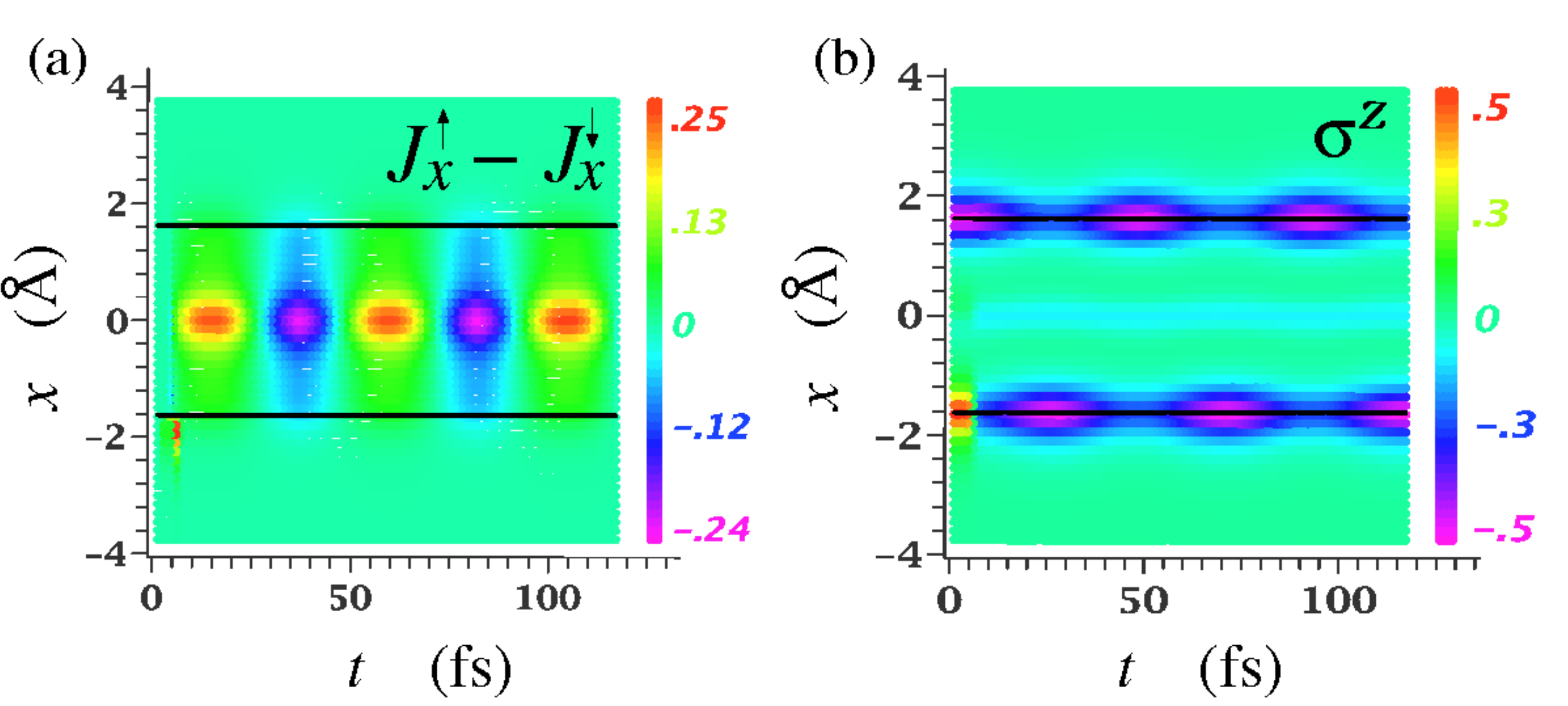}{(Color online) These two graphs are completely analogous to those in Fig. \ref{fig02} but here for the case of H-He-H with $d_\mathrm{H-He}=1.625$~\AA. Note that the exciting magnetic pulse here has been strong enough to nearly reverse the spin of the (leftmost) H atom over which it is applied. As a result, here $\varphi=0.33\pi$ and $\sigma^z$ remains negative.}{fig08}

Similarly to the case of H$_2$, we extract the local spins at the hydrogen sites by using the linear transformation of Eq.~(\ref{lintransf}). Also in this case the transformation seems to work well since it produces consistent results and integration-volume independent local spin (expectation value) magnitudes, angles $\varphi$ and corresponding $J_\mathrm{dyn}$. The variation of electron density along the bond axis as a function of $\varphi$ obtained in the long-time limit is shown in Fig.~\ref{fig09}. This represents a direct density-level signature of the superexchange mechanism. As the spin state goes from LS to HS the charge density at the He atom splits spatially and the two He electrons show a tendency to pair up with the uncoupled hydrogen electrons in the interstitial regions.

\myFig{1}{1}{true}{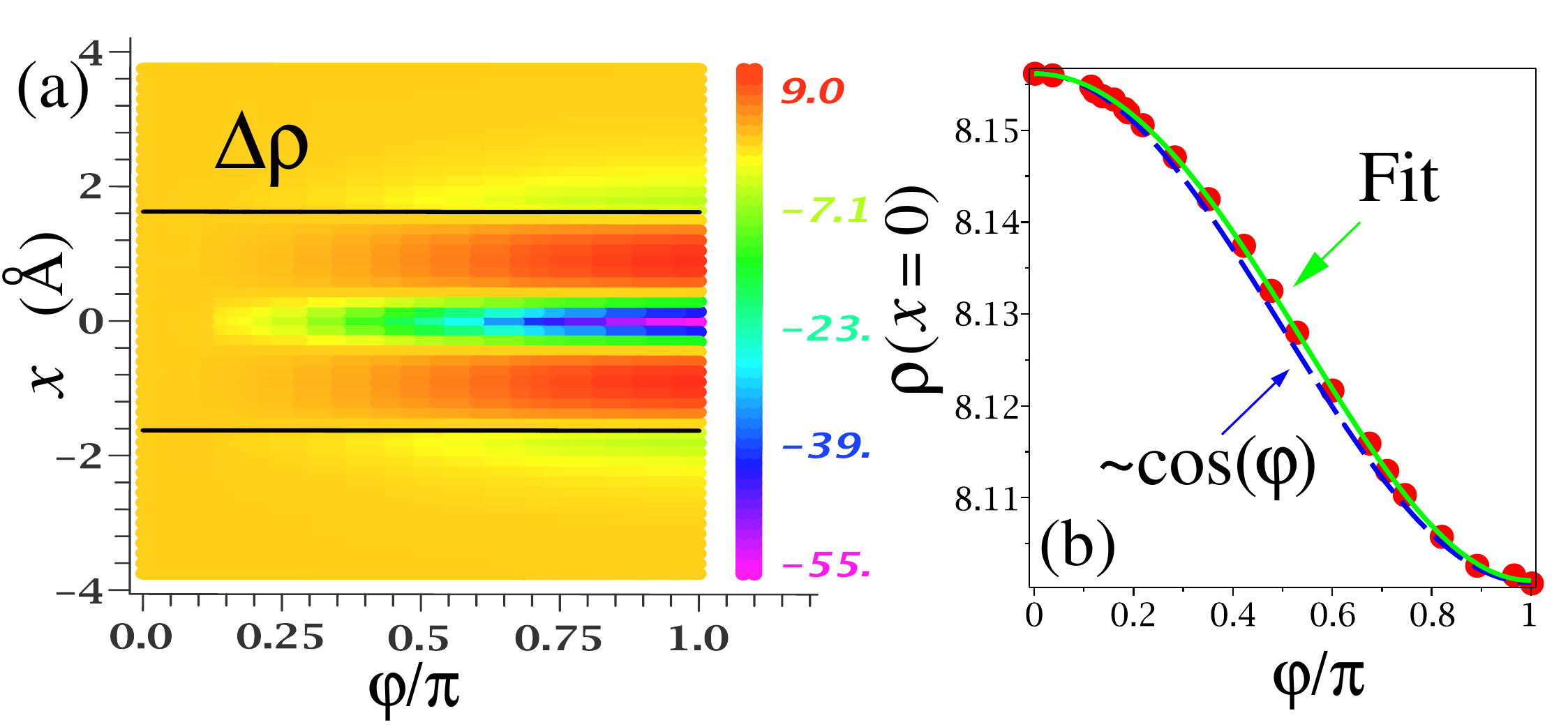}{(Color online) Exactly the same graphs and units as in Fig.\ref{fig05} but for H-He-H with $d_\mathrm{H-He}=1.625$\AA.}{fig09}

The profile of the local density variation with $\varphi$ is again an approximate cosine (plotted in Fig.~\ref{fig09}(b) for the symmetry center but also true everywhere else). In this case the deviation from a perfect cosine dependence is much less pronounced than in the case of H$_2$ (although we find again some higher order harmonic contributions). The better agreement to a cosine function can be attributed both to the variation of the H-H distance in the two cases (2.6~\AA\ for H$_2$ and 3.2~\AA\ for H-He-H) and to the contribution of the superexchange spin-spin coupling mechanism \cite{Yosida}. Similarly, the profile of $E_\mathrm{tot}(\varphi)$ fits to $\cos^2\varphi$ better than in the case of H$_2$. In fact, by using only one additional harmonic to the fitting function, namely $A\cos(\varphi)+B\cos^2(\varphi)$, we find $J_\mathrm{E}$ [see Eq.~(\ref{JE})] to agree extremely well with the extrapolated values of $J_\mathrm{dyn}$ at $\varphi=0$ and $\varphi=\pi$ (see Fig. \ref{fig10}). The relative variation of both the $J$'s ($J_\mathrm{E}$ and $J_\mathrm{dyn}$) calculated either near the LS state or the HS one is also much smaller that the one found in H$_2$. This improvement is to a great extend due to the increased charge localization with inter-hydrogen distance of $d_\mathrm{H-H}=3.3$~\AA\ in H-He-H with respect to 2.6~\AA\ for the H$_2$ [see Fig.~\ref{fig06}(b)]. In fact, for the same small $d_\mathrm{H-H}=2.6$\,\AA, the exchange interaction in H-He-H shows much more substantial deviation from the Heisenberg model (see the Appendix). For the case of the larger distance of 3.3\,\AA, the variation of $J_\mathrm{dyn}$ between the HS and the LS spin is the same in sign and comparable in magnitude to the constained-spin DFT result of Peralta and Barone \cite{Peralta}. They further suggest that a significant portion of that variation is related to the LSDA approximation as $J_\mathrm{E}(\pi)-J_\mathrm{E}(0)$ can be reduced from about 6\% of the average value to less than 1\% with the use of a hybrid XC functional like B3LYP. 

\myFig{1}{1}{true}{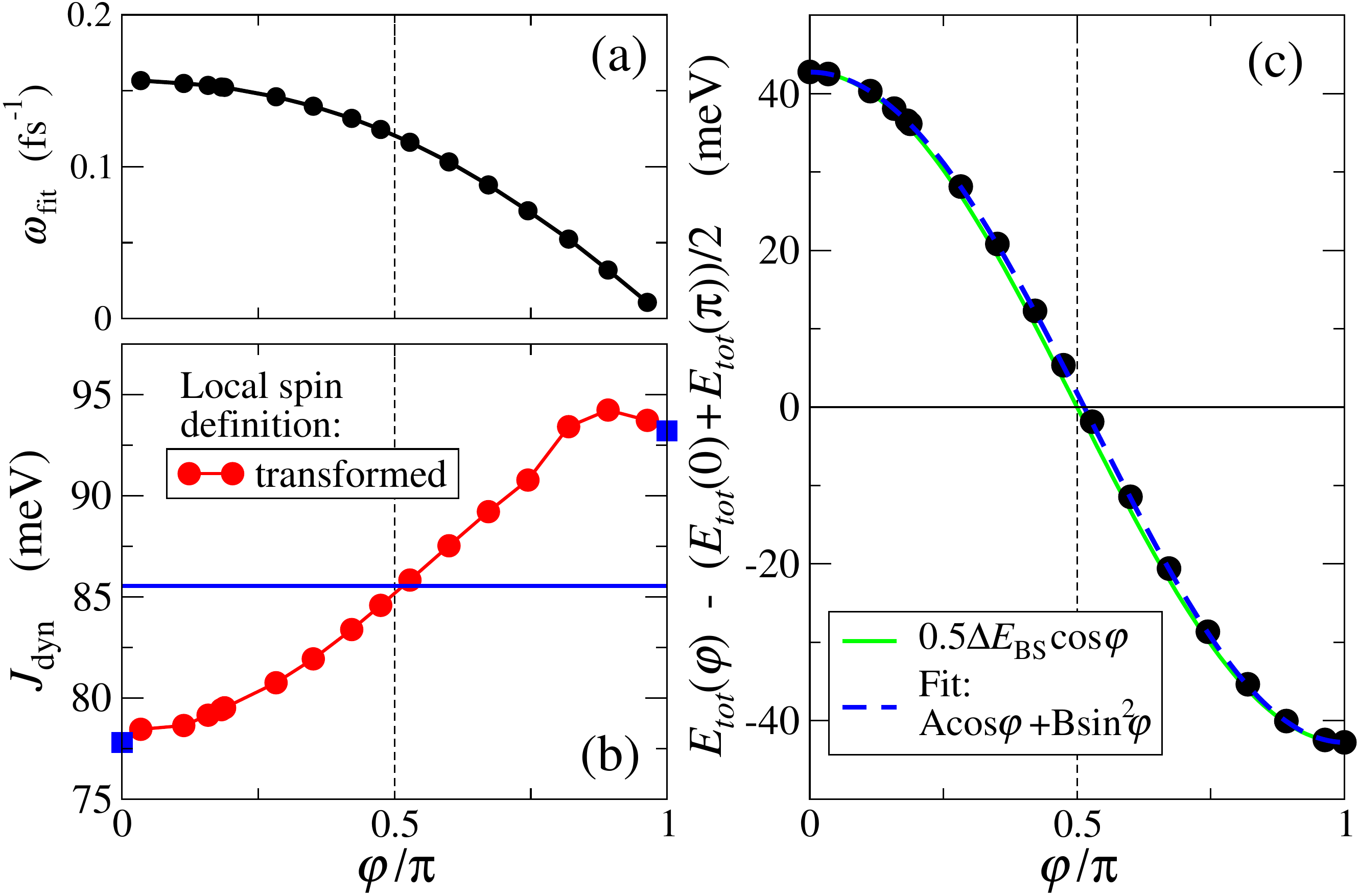}{(Color online) Same graphs as in Fig. \ref{fig06} but for H-He-H with $d_\mathrm{H-He}=1.625$~\AA. The broken symmetry energy difference marked in panel (b) is $\Delta E_\mathrm{BS} = E_\mathrm{HS}-E_\mathrm{LS}=85.5$\,meV. The blue squares correspond to $J_\mathrm{E}$ from Eq. (\ref{JE}) with the fitting function in panel (c), the blue broken line.}{fig10}

We finally present results for the dependence of $J_\mathrm{dyn}$ on the He-H distance, $d_\mathrm{H-He}$. In Fig.~\ref{fig11} the exchange parameter calculated either from the HS or the LS state are compared to broken symmetry DFT~\cite{Bencini,Ruiz} and the exact configuration interaction results available in literature~\cite{Hart}. Figure~\ref{fig11} reveals that $J_\mathrm{dyn}$ converges towards our LSDA broken symmetry value as $d_\mathrm{H-He}$ gets larger. Such convergence in not found for the case of H$_2$ (see Fig.~\ref{fig07}). This means that our dynamical measure of the superexchange interaction in H-He-H suggests that this is much more Heisenberg-like than the direct exchange operating in H$_2$.

\myFig{1}{1}{true}{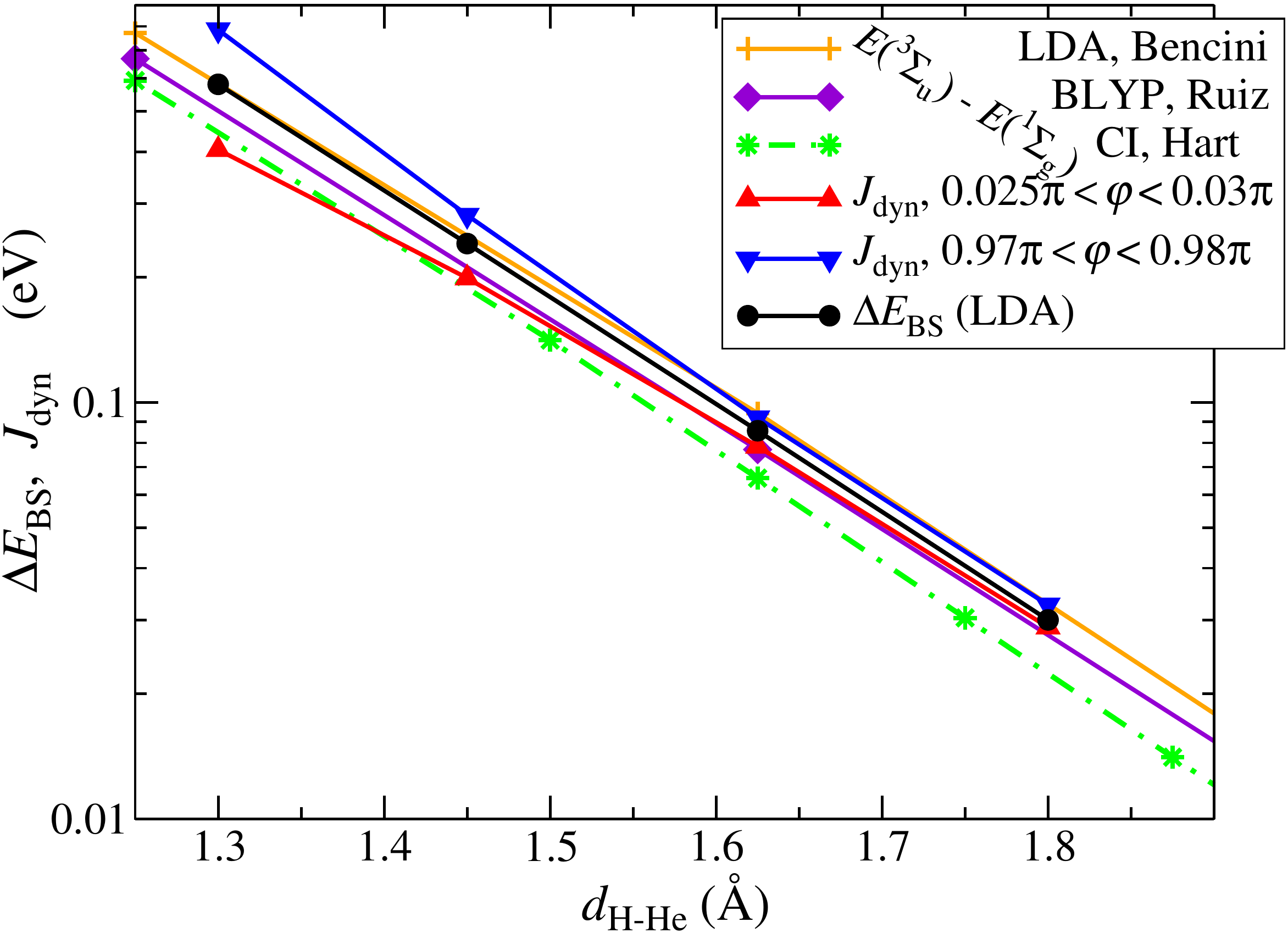}{(Color online) Distance dependence of the exchange parameters for H-He-H calculated by various methods. Here Ruiz, Hart and Bencini correspond respectively to references \onlinecite{Ruiz}, \onlinecite{Hart} and \onlinecite{Bencini}.}{fig11}

\section{Conclusions}

We have demonstrated that the spin dynamics of two simple spin-dimers, as calculated on the basis of TDDFT within the adiabatic LSDA, is rather simple and understandable through a classical model. A non-collinear spin state can be created with inhomogeneous magnetic field pulses and this retains the non-collinearity in the long-time limit. The long-time spin dynamics is thus a harmonic precession in which the non-collinear spin-density rigidly revolves about the total spin of the dimer and all the relative angles remain constant in time. Hence, the trajectories of the localised atomic-like spins, independently from their particular definition, map well onto the classical Heisenberg model. In order for this mapping to be used for the extraction of Heisenberg exchange parameters, the actual definition of local spins is important. 

We have showed how a linear transformation, based on the HS and the LS collinear states and the direct integration of spin density over atomically-centered spheres, can be used to extract the directions of two localized spins. When defined in this way the latter are, to a good degree, independent of the integration sphere used for their definition. This also remains valid for the corresponding dynamically-defined exchange paremeter $J_\mathrm{dyn}$, for a range of distances where the overlap of the atomic wave-function is significant. Such defined exchange parameters agree well with the results from constrained DFT around the LS and the HS states and with broken symmetry total LSDA energy results. We do acknowledge that the actual form of the exchange parameter depends on the choice of exchange and correlation functional used and that our dynamical method does not remedy the shortfall of local and semi-local functionals. We believe that the dynamical method highlighted in this paper, together with generating a quantitatively relevant estimate of $J$, could potentially provide a straightforward verification for the applicability of the Heisenberg spin model to any spin-polarized nano-scaled system. Furthermore, it offers a possible strategy for mapping {\it ab initio} simulations on the widely used atomistic Landau-Lifshitz-Gilbert micromagnetic models for spin-dynamics.

\begin{acknowledgements}
This work has been sponsored by the the European Union under the Cronos project (No. 280879). The authors wish to acknowledge the SFI/HEA funded Irish Centre for High-End Computing (ICHEC) for the provision of computational facilities and support.
\end{acknowledgements}

\begin{appendix}

\section{Dependence of the calculated $J$'\lowercase{s} on the local spin definition: the H-H\lowercase{e}-H case}

We elaborate here on the procedure for extracting the local spins and the exchange parameters for the H-He-H molecule. The leading exchange mechanism in this system is not the direct one, i.e. it does not necessarily depend on the degree of direct overlap of the atomic orbitals at the two magnetic sites. It is then not clear {\it a priori} whether the linear transformation used to eliminate the dependence on the wave-function overlap in H$_2$ is transferable to this case. Indeed, for H-He-H the spin-density snapshots in the long-time limit show a complex texture with multiple peaks and valleys around the He site and the interstitial regions (see Fig.~\ref{fig09}). The main approximation, subsumed in the linear transformation, that the HS and LS state have approximately the same single-electron density distributions (but not spin direction), seems likely to be violated if one looks at the transformation of the spin-distribution between the HS and the LS state as cartooned in Fig.~\ref{fig12}(a). This, however, is not the case and we find that the average variation between the actual density distributions of the LS and HS collinear states at any point in the simulation box is below 5\% for $d_\mathrm{H-He}=1.3$~\AA. With this result at hand, we verify numerically that the linear transformation, described in Section \ref{secLinT}, is still an adequate choice for the local spin definition even at relatively small interatomic distances.

\myFig{1}{1}{true}{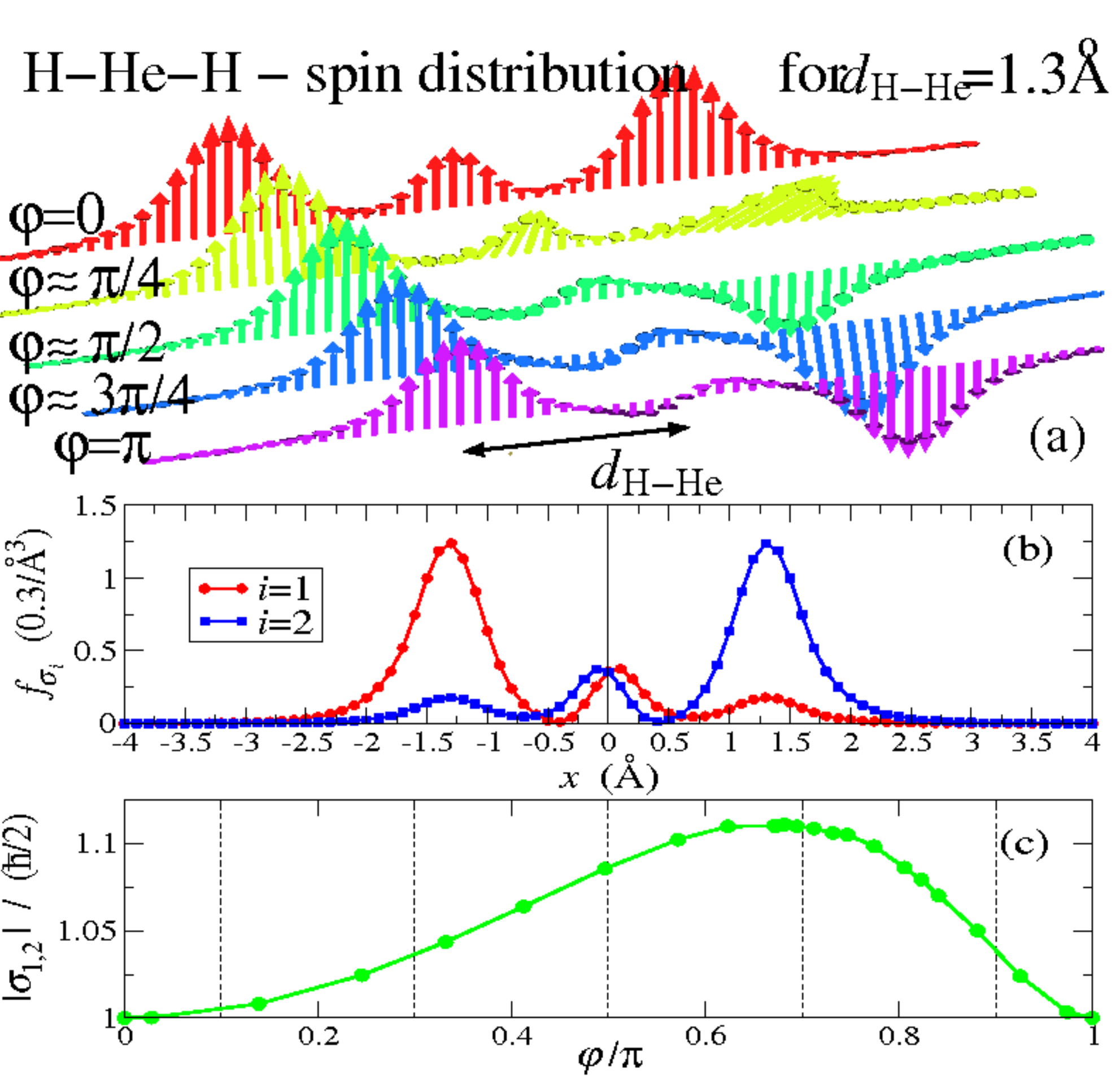}{(Color online) (a) Snapshots of the long-time spin-density along the axis of H-He-H for $d_\mathrm{H-He}=1.3$~\AA\ (note that individual cartoons are rigidly rotated so that the leftmost spins in all snapshots are parallel to each other). (b) The carrier functions of the two spin-density distributions $\vcr{\sigma}_{1}(\vcr{r})$ and $\vcr{\sigma}_{2}(\vcr{r})$ [see Eq. (\ref{sigma_r})], based on the HS and the LS collinear states. (c) Variation of the magnitude of the local spin, defined as $\Abs{\vcr{\sigma}_i} \equiv\Abs{C_i}\int\sigma_i(x)$ (which is identical for $i=1,2$ because of the symmetry), as function of the angle $\varphi$ when the linear combination $\vcr{C_1} \sigma_1(x) + \vcr{C_2} \sigma_2(x)$ is used to fit the non-collinear distributions in panel (a). This result matches exactly $\Abs{\vcr{\sigma}_{1,2}}$ obtained through the linear transformation of Eq. (\ref{lintransf}). }{fig12}

Our first criterion for assessing the adequateness of the local spin definition is the fact that the spins values obtained through the linear transformation do not depend on the choice of the sampling spatial volume, e.g. on the radius, $r_\mathrm{sph}$, of the sphere used to integrate the spin density. We find numerical evidences that this criterion is fulfilled even for small H-He distances where the overlaps are significant. In the top panels of Fig.~\ref{fig13} we compare the total energy dispersion as a function of the angle, $\varphi$, between the two hydrogen local spins, for $\varphi$ determined directly from the apparent spins in an extremely small sphere ($r_\mathrm{sph}=0.05 d_\mathrm{H-He}$), in an extremely large sphere ($r_\mathrm{sph}=d_\mathrm{H-He}$), and the case of $\varphi$ determined after the linear transformation (the blue squares in the graphs). For instance, in the more problematic case of small separation $d_\mathrm{H-He}=1.3$~\AA, the average relative variation in the calculated $\varphi$ (after the linear transformation) is bellow 0.5\% for a variation of $r_\mathrm{sph}$ between $0.05 d_\mathrm{H-He}$ and $d_\mathrm{H-He}$. This constitutes a tiny horizontal error-bar of the blue square data-points in Fig. \ref{fig13}(a), smaller than the symbol size and clearly insignificant on the background of the sphere-radius variation of the apparent local-spin definition. In the case of a large bond-length all the $E_\mathrm{tot}(\varphi)$ curves (for different local sphere definitions) collapse onto one, which tends towards the ideal Heisenberg cosine law [see Fig. \ref{fig13}(b)].

\myFig{1}{1}{true}{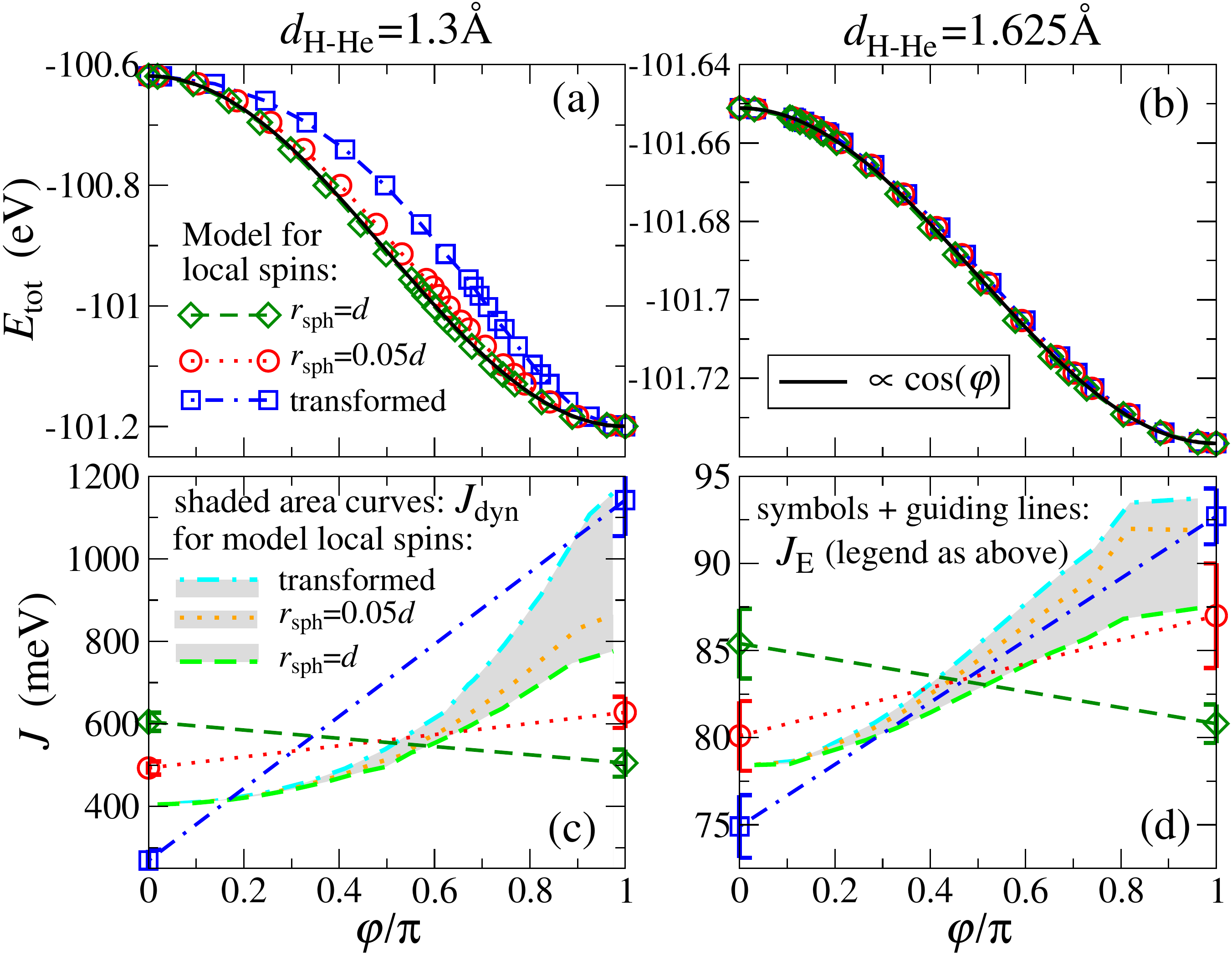}{(Color online) Comparison to an ideal Heisenberg cosine law (solid black curve) of $E_\mathrm{tot}(\varphi)$ (total TDDFT energy in the long-time limit) profiles with angles $\CrB{\varphi}$ corresponding to different definitions of the local spins for two different bond-lengths (a) $d=1.3$~\AA\ and (b) $d=1.625$~\AA. Corresponding $J_\mathrm{E}$ values at the two collinear-spin limits [see Eq. (\ref{JE})] for three different definitions of the local spins, are compared to the dynamical results for $J_\mathrm{dyn}$ (the area shaded in gray). The straight lines are just guides to the eye between the two values. The type of line is matched to the corresponding dynamical result (in or at the border of the gray-shaded region) for the same definition of $\varphi$.}{fig13}

A second relevant criterion could be how well the definition preserves the magnitude of the local spin in the various non-collinear states. Ideally, if the spin-density distributions $\vcr{\sigma}_i(x)$ [Fig. \ref{fig12}(b)], determined from the sum and the difference of spin-density between the LS and HS collinear states, are preserved in the non-collinear state (they only rotate), the linear transformation in Eq. (\ref{lintransf}) will not affect the spin magnitudes in the non-collinear states ($\vcr{\sigma}_i$ are normalized by definition). The result of the linear transformation for $d=1.3$~\AA\ is presented in Fig. \ref{fig12}(c). Clearly, the variations from the norm of 1 are relatively small with a peak at about $\varphi=2/3\pi$, where the local spin is about 11\% larger than its value at $\varphi=0$ or $\varphi=\pi$.

As a final criterion, we consider how well the exchange coupling $J_\mathrm{E}$ extracted from the total energy, Eq.~(\ref{JE}), agrees with the dynamical exchange $J_\mathrm{dyn}$ defined in Eq.~(\ref{Jdyn}). This comparison is presented in the bottom panels of Fig.~\ref {fig13}. Here we also take into an account the fact that different local spin definitions result in different values of the angle $\varphi$. The magnitude of the local spins in Eq.~(\ref{Jdyn}) assumed to be always $\cool{S}=1/2$. When analyzing such a direct comparison we need to keep in mind that the values of $J_\mathrm{E}$ are associated with substantial inaccuracy, as they rely on a numerical second derivative. The error-bars represent the standard deviation of a set of results (of about 20 entries) obtained by using either different form of local interpolation (polynomial) around the end points (0 and $\pi$) or global fits of $E_\mathrm{tot}(\varphi)$ to Fourier cosine series of up to ninth order. 

We find that the worst performing definition is the one based on a large sphere. This systematically produces an incorrect slope of $J(\varphi)$ [see Fig.~\ref {fig13}(c,d)]. Reducing the radius of the sphere improves the agreement, particularly for the larger distance. The result of the linear transformation is rather surprising in the small separation case. It significantly corrects the angles and gives rise to a larger variation of $J_\mathrm{dyn}$ between the two collinear limits. This variation is a signature for the unfitness of the Heisenberg model in this case. At the same time, $J_\mathrm{E}$, based on the second derivatives of $E_\mathrm{tot}(\varphi)$ is also showing a similar variation. this suggests that the classical mapping of the TDDFT spin-dynamics seems to capture the same term in the Hamiltonian as the total energy second derivative. Based on this comparison, it is difficult to argue whether the small sphere or the linear transformation is more suitable for the local spin definition in this molecule. However, the comparison allows us, without analyzing microscopic details, to dismiss some definition (the large sphere, in this case) on the basis that it leads to inconsistent results between the dynamical and the total-energy method for evaluating the Heisenberg exchange coupling $J$. 
\end{appendix}

\end{document}